%% file: 432Reconstruction.tex
\documentclass[a4paper,twoside,openany,11pt]{memoir} %openany, openright
\pdfoutput=1

\def\fullheadfoot{0}
\input{PreRamble}

\newcommand{\braces}[1]{\left\lbrace #1 \right\rbrace}

\newcommand{\Tr}[1]{\mathop{\mathrm{Tr}}\!\big[#1\big]\!}

\newcommand{\eminus}{\vcenter{\hbox{\scalebox{0.6}[1]{$ - $}}}}	%Narrow minus signed (for e.g. negative exponents)

\newcommand{\hc}{\; + \; \mathrm{H.c.} \;}
\newcommand{\andeq}{\quad \mathrm{and} \quad}

\newcommand{\dd}{\mathop{}\!\mathrm{d}}
\newcommand{\ud}[2]{\phantom{}^{#1}\phantom{}_{#2}}
\newcommand{\du}[2]{\phantom{}_{#1}\phantom{}^{#2}} 

\newcommand{\transpose}{^{\mathrm{T}}}
\newcommand{\rep}[1]{\mathbf{#1}}

\newcommand{\sumperm}[1]{\!\! \sum_{#1 \, \mathrm{perm}} \!\!}
\newcommand{\cofg}[2]{\mathbf{g}^{(#1)}_{#2}}
\newcommand{\cofq}[2]{\mathbf{q}^{(#1)}_{#2}}
\newcommand{\cofy}[2]{\mathbf{y}^{(#1)}_{#2}}
\newcommand{\coff}[2]{\mathbf{f}^{(#1)}_{#2}}
\newcommand{\cofs}[2]{\mathbf{s}^{(#1)}_{#2}}

\makeatletter
\newcommand{\vast}{\bBigg@{3}}
\makeatother

\newcommand{\sscript}[1]{{\scriptscriptstyle \mathrm{#1}}}
\renewcommand{\L}{\mathcal{L}}
\newcommand{\LL}{\mathrm{L}}
\newcommand{\RR}{\mathrm{R}}

\newcommand{\SU}{\mathrm{SU}}

\newcommand{\bef}{$ \beta $-function\xspace}
\newcommand{\befs}{$ \beta $-functions\xspace}
\newcommand{\msbar}{$ \overline{\text{MS}} $\xspace}
\renewcommand{\quote}[1]{``#1''}

\colorlet{bluegray}{blue!50!black!20!white}

\begin{document}

% % % % Title % % % % 
\thispagestyle{empty}
\renewcommand*{\thefootnote}{\fnsymbol{footnote}}
\vspace*{0.01\textheight}
\begin{center}
	{\sffamily \bfseries \LARGE \mathversion{chaptermath} 
		General Gauge--Yukawa--Quartic $\beta$-Functions at \\[.2em] 4--3--2-Loop Order
	}\\[-.5em]
	\textcolor{blue!80!black}{\rule{.9\textwidth}{2.5pt}}\\
	\vspace{.02\textheight}
	{\sffamily  \mathversion{subsectionmath} \large 
	Joshua Davies,$ ^{1} $\footnote{j.o.davies@sussex.ac.uk} 
	Florian Herren,$ ^{2} $\footnote{florian.s.herren@gmail.com} 
	and Anders Eller Thomsen$ ^{3} $\footnote{anders.thomsen@unibe.ch}}
	\\~
	\\{ \small \sffamily \mathversion{subsectionmath} 
	$ ^{1}\, $Department of Physics and Astronomy, University of Sussex, \\[-.3em] Brighton, BN1 9HQ, UK 
	\linebreak $ ^{2}\, $Fermi National Accelerator Laboratory, \\[-.3em] Batavia, IL, 60510, USA
	\linebreak $ ^{3}\, $Albert Einstein Center for Fundamental Physics, Institute for Theoretical Physics,\\[-.3em]
	University of Bern, CH-3012 Bern, Switzerland
	}
	\\[.005\textheight]{\itshape \sffamily \today}
	\\[.03\textheight]
\end{center}
\setcounter{footnote}{0}
\renewcommand*{\thefootnote}{\arabic{footnote}}%
\suppressfloats	%Prevents figures and the likes on this page (title page)
% % % % Title % % % % 

\begin{abstract}\vspace{-.05\textheight}
We determine the full set of coefficients for the completely general 4-loop gauge and 3-loop Yukawa $ \beta $-functions for the most general renormalizable four-dimensional theories. Using a complete parametrization of the $ \beta $-functions, we compare the general form to the specific $ \beta $-functions of known theories to constrain the unknown coefficients. The Weyl consistency conditions provide additional constraints, completing the determination.  \\		
{\vspace{.7em} \footnotesize \itshape Preprint: FERMILAB-PUB-21-471-T}
\end{abstract}

\toc

\section{Introduction}
The Renormalization Group (RG) is a key concept in quantum field theory, which determines the change in physics due to changing energy scales. Among numerous applications, its role in improving perturbative computations makes it indispensable for precision computations both in and beyond the Standard Model (SM).
Completely general results for the 2-loop \befs of gauge-Yukawa theories have served the community well for almost four decades~\cite{Machacek:1983tz,Machacek:1983fi,Machacek:1984zw,Jack:1984vj}, with the noteworthy addition of the 3-loop contribution for simple gauge groups~\cite{Pickering:2001aq}. 

In the last decade, the discovery of the Higgs and a new drive for precision physics prompted the calculation of the SM \befs to 4-loop order for gauge and 3-loop order for Yukawa and quartic couplings~\cite{Mihaila:2012fm,Chetyrkin:2013wya,Bednyakov:2013cpa,Bednyakov:2014pia,Bednyakov:2015ooa,Zoller:2015tha,Davies:2019onf}.
Recent years have also seen new developments in hitherto neglected aspects of \befs, namely, the \befs for the relevant operators~\cite{Luo:2002ti,Schienbein:2018fsw,Sartore:2020pkk}, the running of the kinetic-mixing parameters between Abelian gauge fields~\cite{Luo:2002iq,Fonseca:2013bua,Poole:2019kcm}, and the extension of the 3-loop gauge \bef to semi-simple gauge groups~\cite{Mihaila:2014caa,Poole:2019kcm}.
All these developments have been collated in a number of accessible computer tools---\texttt{SARAH 4}~\cite{Staub:2008uz,Staub:2013tta}, \texttt{PyR@TE 3}~\cite{Sartore:2020gou,Lyonnet:2013dna}, \texttt{ARGES}~\cite{Litim:2020jvl}, and \texttt{RGBeta}~\cite{Thomsen:2021ncy}---for the application of the general formulas to specific theories, which have opened the doors for the wider application of higher-loop general results in the community.  
The bespoke tool \texttt{RGE++}~\cite{Deppisch:2020aoj} has also been made available for fast evaluation of the resulting RG equations.  

New insights into the underlying structure of the RG have been obtained in recent years on the basis of the powerful machinery of the \emph{local RG} (LRG)~\cite{Shore:1986hk,Osborn:1989td,Jack:1990eb,Osborn:1991gm,Jack:2013sha,Baume:2014rla}. 
Crucially, as it pertains to the present matter, it was pointed out that Osborn's equation implies nontrivial consistency conditions on the \bef, now known as \emph{Weyl consistency conditions} (WCC). 
The WCCs have since been leveraged to check the self-consistency of already-known \befs and to make predictions for as-of-yet undetermined \bef coefficients~\cite{Antipin:2013sga,Jack:2013sha,Jack:2014pua,Gracey:2015fia,Poole:2019kcm,Poole:2019txl}. 
These endeavors have emphasized the usefulness of parametrizing the unknown \befs in terms of a basis of tensor structures (TS) constructed from the couplings of the theory in order to attempt to fix the coefficients of the parametrization. 
By comparing the generic expression to known results in specific theories, the 3-loop \befs for pure Yukawa theories were almost entirely fixed without the need for any new, arduous loop computations~\cite{Steudtner:2020tzo,Steudtner:2021fzs}.     

This paper represents the culmination of several years of research by finally fixing the full gauge-Yukawa-quartic \befs to 4--3--2-loop order.
To this end, we use the TS basis of Ref.~\cite{Poole:2019kcm} along with the WCCs derived therein. 
We fix the remaining coefficients by matching the generic expressions to the known \befs of the 4-loop SM gauge~\cite{Davies:2019onf}, the 3-loop 2HDM Yukawa~\cite{Herren:2017uxn}, and several new SM-adjacent results computed with the express purpose of fixing the remaining coefficients.   

While we were finalizing our computations, another paper with overlapping scope appeared~\cite{Bednyakov:2021qxa}.
The two computations agree and serve as valuable crosschecks. 
However, both computations rely on the \bef parametrization of Ref.~\cite{Poole:2019kcm} and the tensor implementation of \texttt{RGBeta}~\cite{Thomsen:2021ncy}, making for a single point of failure. 
That said, such a failure appears highly unlikely due to the required self-consistency of both computations. 

The remainder of the paper is organized as follows: In the next section, we review the necessary aspects of the LRG and introduce the \bef parametrization, as well as our computational setup.
In section~\ref{sec::models}, we describe the models used to fix the remaining coefficients and discuss the combination of the various models. Finally, we conclude in section~\ref{sec::conc}.
We provide explicit expressions for the TSs of the gauge and Yukawa \befs in the appendix along with our results for all coefficients.

\section{Setup and Methods}

\subsection{Notation and conventions}
In the mass-independent \msbar scheme, the \befs of marginal couplings are independent of the relevant couplings. Accordingly, we restrict our attention to the most general four-dimensional theory allowing only for marginal couplings:  
	\begin{align} \label{eq:gen_Lagrangian}
	\L = &- \tfrac{1}{4} a^{\eminus 1}_{AB} F^{A}_{\mu\nu} F^{B\mu\nu}  + \tfrac{1}{2} (D_\mu \phi)_{a} (D^{\mu} \phi)_a + i \psi_i^{\dagger} \bar{\sigma}^\mu (D_\mu \psi)^i \nonumber \\
	&- \tfrac{1}{2} \left(Y_{aij} \psi^i \psi^j \hc \right) \phi_a - \tfrac{1}{24} \lambda_{abcd} \phi_a \phi_b \phi_c \phi_d.
	\end{align}
To be clear, when we name this the \quote{most general} theory, it is because any other renormalizable theory can be cast in this exact form.\footnote{In practice recasting theories in this form can be a rather grueling task. Methods to do it systematically have been developed, see e.g. Ref.~\cite{Molgaard:2014hpa}.} 
In this construction, all scalar fields are collected into the real scalar multiplet $ \phi_a $, while the fermions, allowing for chiral theories, are collected in the multiplet $ \psi^{i} $ using Weyl spinor notation. 
Thus, we avoid having to distinguish between left- and right-handed fields. 

In this framework $ \phi_a $ and $ \psi^i $ are generically in reducible representations of the gauge group $ G $. 
The covariant derivatives of these fields are given by 
	\begin{equation}
	\begin{split}
	D_\mu \phi_a &= \partial_\mu \phi_a - i A_{\mu}^{A} (T^{A}_{\phi})_{ab} \phi_b\,, \\
	D_\mu \psi^{i} &= \partial_\mu \psi^{i} - i A_{\mu}^{A} (T^{A}_{\psi})\ud{i}{j} \psi^{j}\,.
	\end{split}
	\end{equation}
As we allow for any gauge group, it is convenient to move the gauge couplings of all the simple product groups and Abelian factors to the kinetic term, where they are incorporated in the gauge coupling matrix $ a_{AB} $~\cite{Poole:2019kcm}.

It is convenient to arrange the fermion index space in a Majorana-like manner to incorporate both left and right chiralities simultaneously~\cite{Jack:2014pua,Poole:2019kcm}. The Yukawa couplings and gauge generators are then given by 
	\begin{equation} 
	y_a = \begin{pmatrix}    Y_a & 0 \\ 0 & Y_{a}^{\ast } \end{pmatrix} \andeq T^{A} = \begin{pmatrix} T_{\psi}^{A} & 0 \\ 0 & -T_{\psi}^{A\,\ast} \end{pmatrix},  
	\end{equation} 
and we define the chirality-flipped derivative quantities by 
	\begin{equation}
	\tilde{y}_a = \sigma_1 y_a \sigma_1 = \begin{pmatrix}    Y_a^\ast & 0 \\ 0 & Y_{a} \end{pmatrix} \andeq \widetilde{T}^{A}= \sigma_1 T^A \sigma_1 = \begin{pmatrix} - T_{\psi}^{A,\, \ast} & 0 \\ 0 & T_{\psi}^{A} \end{pmatrix},
	\end{equation} 	
with a similar construction for all other ``tilde'' quantities.\footnote{In the context of the fermion indices, the $\sigma$ matrices are to be understood as a tensor product of Pauli matrices with the identity matrix.}
In this notation, the $ \gamma_5 $-odd contribution to the 4-loop gauge and 3-loop Yukawa are characterized by the appearance of $ \sigma_3 $ on the fermion lines: 
the underlying diagrams carry opposite signs for the different chiralities. 

The perturbative \befs of the tensor couplings are expanded by loop orders and parametrized as
    \begin{align} \label{eq:beta_parametrization}
	\beta_{AB} &= \dfrac{\dd a_{AB}}{\dd t} = \sum_{\ell=1}^\infty a_{AC} \dfrac{ \beta^{(\ell)}_{CD} }{(4 \pi)^{2\ell}} a_{DB}\,, \\
	\beta_{aij} &= \dfrac{\dd y_{aij}}{\dd t}  = \sum_{\ell=1}^\infty \dfrac{\beta^{(\ell)}_{aij}}{(4\pi)^{2\ell}}\,, \\
	\beta_{abcd} &= \dfrac{\dd \lambda_{abcd}}{\dd t}  = \sum_{\ell=1}^\infty \dfrac{\beta^{ (\ell) }_{abcd} }{(4 \pi)^{2 \ell}}\,,
	\end{align}
where $ t=\ln \mu $ is the RG time. 
These are the main objects of our study.

\subsection{Tensor parametrization}
As all other perturbative quantities, the \befs are polynomials in the couplings, of an order determined by the loop order.
It follows that the $ \ell $-loop contributions to the \befs can be parametrized by
	\begin{align}
	\beta^{(\ell)}_{AB} &= \sum_n \cofg{\ell}{n} \, [G^{(\ell)}_{n}(a,\, y,\, \lambda,\, T,\, T_\phi, F)]_{AB}\,, \label{eq:beta_gauge_l}\\	
	\beta^{(\ell)}_{aij} &= \dfrac{1}{2} \, \sumperm{2} \sum_n \cofy{\ell}{n} \, [Y^{(\ell)}_{n}(a,\, y,\, \lambda,\, T,\, T_\phi, F)]_{aij}\,, \label{eq:beta_yukawa_l}\\
	\beta^{(\ell)}_{abcd} &= \dfrac{1}{24} \, \sumperm{24} \sum_n \cofq{\ell}{n} \, [Q^{(\ell)}_{n}(a,\, y,\, \lambda,\, T,\, T_\phi, F)]_{abcd}\,. \label{eq:beta_quartic_l}
	\end{align}
The coefficients $ \cofg{\ell}{n} $, $ \cofy{\ell}{n} $, and $ \cofq{\ell}{n} $ are real numbers, while $ [G^{(\ell)}_{n}]_{AB} $, $ [Y^{(\ell)}_{n}]_{aij} $, and $ [Q^{(\ell)}_{n}]_{abcd} $ are tensors constructed from a combination of couplings, gauge generators, and structure constants ($ F $), referred to as \emph{tensor structures} (TSs). 
By explicitly symmetrizing the Yukawa and quartic \befs---hence the sums over all index permutations---the TSs are kept as simple as possible; they are monomials rather than polynomials.

Whereas a determination of $ \beta^{(\ell)} $ typically requires a full perturbative loop computation, it is possible to find a suitable basis of TSs from general considerations, thereby leaving only the coefficients unknown. 
The main obstacle to finding such a basis is the sheer number of tensors at higher orders and the need to eliminate the redundancy in the parametrization due to the gauge identities stemming from all of the couplings being singlets under the symmetry. 
One suitable basis was determined in Ref.~\cite{Poole:2019kcm}, and we will employ it for our computation here too.
We have taken the liberty of presenting the full basis in regular tensor notation in the Appendix for posterity, as they have only been made available graphically in Ref.~\cite{Poole:2019kcm}. 

We implemented the TSs of the 4-loop gauge and 3-loop Yukawa \befs in \texttt{RGBeta}, which allows us to evaluate Eqs.~(\ref{eq:beta_gauge_l}--\ref{eq:beta_quartic_l}) in specific models and match the generic expressions to the \befs determined with traditional loop computations.

\subsection{Flavor-improved $ \beta $-function} \label{sec:flavor-improvement}
The \msbar \befs are generally not unique beyond the second loop order. The flow in the direction of flavor symmetry transformations (the symmetry of the kinetic term of Lagrangian~\eqref{eq:gen_Lagrangian}) is fixed by the choice of renormalization constants, which are themselves ambiguous~\cite{Herren:2021yur}. 
It can even be divergent in the dimensional expansion, although physics is insensitive to any such flavor rotations along the flow due to the symmetry of the action under combined field and coupling transformations. 
In one instance, a perturbative limit cycle was found but soon ascribed to a non-physical flow along the flavor group transformation, the physics being that of a regular fixed point~\cite{Fortin:2012cq,Fortin:2012hn}.

The flavor-improved \bef $ B_I $ is a particular construction of the \bef engineered to be invariant under the ambiguity~\cite{Fortin:2012hn,Jack:2013sha,Baume:2014rla} and has been shown to be independent of the choice of renormalization constants~\cite{Herren:2021yur}. 
It factors out the flow along the flavor-symmetric direction, thereby avoiding the appearance of non-physical limit cycles in the RG flow. It is given by 
	\begin{equation}
	B_I = \beta_I - (\upsilon \, g)_I,
	\end{equation}
where $ I,J,\ldots $ are coupling indices taking values of all the couplings of the theory $ g_I = \braces{a_{AB},\, y_{aij},\, \lambda_{abcd}} $. $ \upsilon $ is another RG function of the LRG associated with the current of the flavor symmetry group.\footnote{Inconveniently, $ \upsilon $ (lower-case upsilon) is referred to as $ S $ in parts of the literature.} 
A conformal field theory is characterized by $ B_I =0 $ and not, as is often said, $ \beta_I=0 $. 
That being said, in simple/simplified theories in which the couplings are trivial w.r.t.~the flavor group, the two functions coincide.   

In Ref~\cite{Herren:2021yur}, we found that there always exists a choice of renormalization constants for which $ \beta_I $ is identical to $ B_I $ (meaning that $ \upsilon=0 $). 
Thus it is perfectly valid to use $ B_I $ in place of $ \beta_I $ for the RG, a choice that also carries the benefits discussed above. 
Unfortunately, there is no direct prescription for the renormalization constants for which this identification is valid, and $ B_I $ will have to be determined directly from $ \beta_I $ and $ \upsilon $. 

$ \upsilon $ is, potentially, nontrivial starting from $ 3 $-loop order in gauge-Yukawa theories, but only when the flavor structure is nontrivial. 
The flavor symmetry acts trivially on the gauge group, so $ (\upsilon \, a)_{AB} =0 $. Thus, it is only the 3-loop Yukawa \bef that is different from its flavor-improved cousin at 4--3--2-loop order. 
The action of $ \upsilon $ on the Yukawa coupling is given by
	\begin{equation} \label{eq:ups_action}
	(\upsilon\, y)_a = \upsilon_{ab}\, y_b + \upsilon\, y_a - y_a \tilde{\upsilon}\,.
	\end{equation}
$ \upsilon $ is technically a single element of the Lie algebra of the flavor group. 
The scalar and fermion indices on $ \upsilon_{ab} $ and $ \upsilon\du{i}{j} $ (as per usual we use implicit fermion indices in Eq.~\eqref{eq:ups_action}) are implicitly taken to refer to different representations of $ \upsilon $ as determined by the transformation of the scalar and fermion fields.
We parametrize $ \upsilon $ at $ \ell $-loop order by 
	\begin{align}
	\upsilon^{(\ell)}\du{i}{j} &= \sum_n \coff{\ell}{n} [F^{(\ell)}_n - \widetilde{F}^{(\ell)\, \mathrm{T}}_n]\du{i}{j}\,, \\
	\upsilon^{(\ell)}_{ab} &= \sum_n \cofs{\ell}{n} \big([S^{(\ell)}_n]_{ab} - [S^{(\ell)}_n]_{ba}\big)\,,
	\end{align}
with the TSs $ F_n^{(\ell)} $ and $ S_n^{(\ell)} $. A basis for this parametrization was provided in~\cite{Poole:2019kcm}.

\subsection{Weyl consistency conditions} \label{sec:WCCs}
Using the LRG, Jack and Osborn~\cite{Osborn:1989td,Jack:1990eb,Osborn:1991gm,Jack:2013sha,Baume:2014rla} demonstrated a perturbative $ A $-theorem via Osborn's equation
	\begin{equation} \label{eq:Oeq}
	\partial^I \! A = T^{IJ} B_J\,,
	\end{equation}
where $ \partial^I \equiv \partial / \partial g_I $. 
Osborn's equation relates the function $ A $, which coincides with the Weyl anomaly coefficient of the Euler density at fixed points, with the flavor-improved \bef via $ T^{IJ} $, which itself is a combination of various coefficients of the Weyl anomaly. 
Since $ T^{IJ} $ is positive-definite at leading order in the loop expansion, this establishes a perturbative $ A $-theorem. 

As it transpires, Osborn's equation implies nontrivial self-consistency conditions on the \befs, known as Weyl consistency conditions (WCCs)~\cite{Antipin:2013sga,Jack:2013sha,Jack:2014pua,Gracey:2015fia}. 
$ A $ and $ T^{IJ} $ can be parametrized in terms of the couplings of the theory and the unknown coefficients eliminated from Eq.~\eqref{eq:Oeq}, leaving a number of nontrivial consistency conditions between the coefficients of the \befs. Ref.~\cite{Poole:2019kcm} extended this program to the generic gauge-Yukawa theory~\eqref{eq:gen_Lagrangian} and up to order 4--3--2 in the \befs. 
With full knowledge of the known 3--2--2 \befs, the resulting WCCs provide 266 independent conditions on the 510 coefficients $ \cofg{4}{n} $ and $ \cofy{3}{n} $ parametrizing the 4-loop gauge and 3-loop Yukawa \befs, respectively. 
In particular, the conditions unambiguously fix the $ \gamma_5 $ ambiguity in $ \beta^{(4)}_{AB} $ on basis of the known $ \gamma_5 $-odd contribution to $ \beta^{(3)}_{aij} $~\cite{Poole:2019txl}, also confirming the prescription in Ref.~\cite{Bednyakov:2015ooa}. 

The coefficients of $ \upsilon $ can be eliminated from Osborn's equations, leaving a set of WCCs exclusively relating \bef coefficients. This elimination fixes all 3 $ \cofs{3}{n} $ coefficients in terms of the \bef coefficients. 
The 6 $ \coff{3}{n} $ coefficients are fixed up to one free parameter.
Meanwhile, the explicit SM evaluation of the pure Yukawa part of $ \upsilon $ in Ref.~\cite{Herren:2021yur} fixes $ \coff{3}{4,5} $; this is enough to fix the remaining parameter and provide an additional consistency check.
Thus, $ \upsilon $ can be fixed once the \bef coefficients are known, allowing us to fix the 3-loop Yukawa $ B_{aij} $.

\subsection{Loop Computations}
To fix the remaining coefficients, gauge and Yukawa \befs need to be computed in various models. To this end, we follow methods outlined in Refs.~\cite{Herren:2017uxn,Davies:2019onf,Herren:2021yur}. In the following,
we present our computational setup and discuss issues arising in such calculations.

Gauge and Yukawa \befs are related to the counterterms of the couplings. Coupling counterterms can be related to renormalization constants of vertices that are proportional to the respective coupling at tree-level, as well as the renormalization constants of the fields involved in the vertex.
As a consequence, 3-loop Yukawa counterterms can be computed from 3-loop renormalization constants of fermion and scalar fields
as well as fermion-scalar vertices, and 4-loop gauge coupling counterterms can be obtained from 4-loop renormalization constants of gauge and ghost fields, and ghost-gauge vertices.

Our well-tested computational system has been introduced in Ref.~\cite{Davies:2019onf} and subsequently improved for the computations in Ref.~\cite{Herren:2021yur}.
It relies on \texttt{QGRAF} \cite{Nogueira:1991ex} for diagram generation, followed by \texttt{q2e} and \texttt{exp} \cite{Harlander:1997zb,Seidensticker:1999bb} for translating Feynman diagrams
into \texttt{FORM} \cite{Ruijl:2017dtg} expressions and mapping them onto integral families, as well as \texttt{COLOR} \cite{vanRitbergen:1998pn} to compute $\SU(N)$ color factors.
Diagrams sharing the same color factors, flavor structures, and integral families are grouped into so-called ``superdiagrams'', which allows for the efficient computation
of $\mathcal{O}(10^6)$ 4-loop Feynman diagrams; cancellations between different diagrams occur in early stages of the evaluation of superdiagrams, rather than in the final summation of all diagrams.
The superdiagrams are processed using \texttt{FORM} and the Feynman integrals are evaluated by \texttt{FORCER} \cite{Ruijl:2017cxj}.
Aside from the computational complexity involved with 4-loop computations, there are two main issues: the treatment of $\gamma_5$ in $D$ dimensions and the ambiguities in Yukawa \befs.

Starting from 3-loop order, nontrivial contributions from $\gamma_5$-odd traces arise in the computation of Yukawa \befs. These contributions can be computed by evaluating such
traces in 4 dimensions, as has been done for both the SM and the various 2HDMs \cite{Bednyakov:2012en,Chetyrkin:2012rz,Herren:2017uxn}. Matching Eq.~\eqref{eq:beta_yukawa_l} to these models allows us to fix the coefficients
of all 5 $\gamma_5$-odd tensors structures. As discussed in Sec.~\ref{sec:WCCs}, the WCCs then allow us to fix the coefficients of the $\gamma_5$-odd tensor structures in the gauge \bef.
Therefore, we ignore all $\gamma_5$-odd contributions in our calculations here and focus on the missing coefficients in the $\gamma_5$-even sector.

The second issue relates to the ambiguity of the 3-loop Yukawa \bef. 
As observed in \cite{Bednyakov:2012en,Herren:2017uxn}, the presence of global symmetries of the kinetic terms of fermion and scalar fields leads to ambiguities in the wave-function renormalization constants, which in turn lead to an ambiguity in the Yukawa \bef. 
As well as being ambiguous, the Yukawa \bef is, in general, not finite. 
As we noted in Section~\ref{sec:flavor-improvement}, this can ultimately be resolved by determining the flavor-improved \bef $ B_I $.
Here we proceed by choosing Hermitian square roots of the wave-function renormalization constants, consistent with the one-particle-irreducible \bef parametrization of Ref.~\cite{Poole:2019kcm}. 
We extract only the finite part of the RG functions from this and rely on it being possible to modify the renormalization constants in such a way as to cancel the divergences without changing the finite part~\cite{Herren:2021yur}.

\section{Fixing the $ \beta $-Functions}\label{sec::models}
\subsection{The models we use as input}
While the \befs of the SM and 2HDMs in combination with the WCCs provide a large number of constraints on the 510 free coefficients in the 4-loop gauge and 3-loop Yukawa \befs, a number of
coefficients of TSs involving scalar fields remain unconstrained. To constrain these, we consider two types of models: one with scalars charged under two non-Abelian gauge groups
and another in which the scalar field is in a nontrivial representation of the gauge group different from that of the fermions.
In the following, we describe two models that possess the required properties. First, we use a leptoquark model, i.e.,~an extension of the SM with a scalar field charged under $\SU(3)$ that couples quarks to leptons.
The one remaining coefficient that can not be fully constrained by combining the leptoquark model, the SM, 2HDMs, and the WCCs can be fixed by a simple toy model; an $\SU(2)$ gauge theory with a vector-like fermion in the fundamental representation
and a scalar in the adjoint representation.

\subsubsection{Leptoquark model}
Let $ \phi \sim (\rep{3}, \rep{2}, 7/6) $ be a scalar leptoquark added to the SM field content. The Yukawa interactions of the model are then given by 
	\begin{equation}
	\L_{\mathrm{yuk}} = \L_{\text{SM,yuk}} - y_1^{ij} \varepsilon^{\alpha \beta} \phi^\ast_{c\beta} \overline{\ell}_{\LL,\alpha}^{i} u_{\RR}^{j,c} - y_2^{ij} \phi^{c\alpha} \overline{q}_{\LL,c\alpha}^{i} e_{\RR}^{j} \,,
	\end{equation}
where $ c,d,\ldots $ are color, $ \alpha, \beta,\ldots $ isospin, and $ i,j $ generation indices. $ \varepsilon^{\alpha \beta} = i\sigma_2 $ is the invariant of $ \SU(2)_\LL $. 
For comparison, in this notation the SM Yukawas are given by 
	\begin{equation}
	\L_{\text{SM,yuk}} = - y_u^{ij}  \varepsilon^{\alpha \beta} H^\ast_{\beta} \overline{q}_{\LL,c\alpha}^{i} u_{\RR}^{j,c} - y_d^{ij}  H^{\alpha} \overline{q}_{\LL,c\alpha}^{i} d_{\RR}^{j,c} - y_e^{ij}  H^{\alpha} \overline{\ell}_{\LL,\alpha}^{i} e_{\RR}^{j}\,.
	\end{equation}

The quartic couplings are given by 
	\begin{equation}
	V = \dfrac{\lambda_1}{2} (H^{\dagger} H)^2 + \lambda_2 (H^{\dagger} H) \Tr{\phi^\dagger \phi} + \dfrac{\lambda_3}{2} \mathrm{Tr}^2\big[\phi^\dagger \phi \big] +\kappa_1 H\transpose \phi^\dagger \phi H^\ast + \dfrac{\kappa_2}{2} \Tr{\phi^\dagger \phi \,\phi^\dagger \phi}\,,
	\end{equation}
where $ \phi^{c\alpha} $ is thought of as a matrix with indices in color and isospin respectively. In index notation, the $ \kappa_1 $ coupling is given by $ \kappa_1 H^{\alpha} \phi^{\ast}_{c\alpha} \phi^{c\beta} H^{\ast}_\beta $.

\subsubsection{$ \SU(2) $ toy model}
Next, we consider a toy model with a gauge group $ G= \SU(2) $, a real scalar triplet $ \phi_a $, and a vector-like fermion doublet $ \psi^\alpha $. 
This theory is described by the Lagrangian
	\begin{equation}
	\L_\sscript{TM} = -\dfrac{1}{4g^2} (F^{A}_{\mu\nu})^2 + \tfrac{1}{2}(D_\mu \phi_a)^2 + i \overline{\psi}_\alpha \slashed D \psi^\alpha - \dfrac{y}{2} \phi_a \overline{\psi}_\alpha  \sigma\ud{a\alpha}{\beta} \psi^\beta -\tfrac{1}{8} \lambda  (\phi_a \phi_a)^2\,, 
	\end{equation}
where $ \alpha,\beta $ are doublet indices and $ a $ is a triplet index of $ G $. $ \sigma^a/2 $ are the generators of the fundamental representation of $ \SU(2) $, where $ \sigma^a $ are the Pauli matrices.
This toy model, with the scalar in a non-fundamental representation, is rather different from the SM-like models we otherwise consider. 
This helps us to resolve the last 1-parameter degeneracy in $ \cofg{4}{n} $ in terms with both gauge and Yukawa couplings.

\subsection{Fixing the \bef coefficients}
\begin{table}
	\centering
	\begin{tabularx}{.7\textwidth}{| Y | Y Y Y Y |}
	\hline \hline 
	Gauge & SM & 2HDM+$ \nu_\RR $ & SM+LQ & TM \\ \hline 
	SM& 147 & 157 & 196 & 153 \\ 
	2HDM+$ \nu_\RR $ & \cellcolor{bluegray} & 150 & 196 & 154\\ 
	SM+LQ & \cellcolor{bluegray} & \cellcolor{bluegray} & 190 & 192 \\
	TM & \cellcolor{bluegray} & \cellcolor{bluegray} & \cellcolor{bluegray} & 10 \\
	\hline \hline 
	\end{tabularx}
	\caption{The number of independent constraints on the gauge coefficients from combining any two of the models, with the diagonal being the constraints from the individual models.}
	\label{tab:gauge_constraints}
\end{table}

\begin{table}
	\centering
	\begin{tabularx}{.55\textwidth}{| Y | Y Y Y|}
	\hline \hline 
	Yukawa & SM & 2HDM & SM+LQ  \\ \hline 
	SM& 145 & 255 & 285 \\ 
	2HDM & \cellcolor{bluegray} & 246 & 292 \\ 
	SM+LQ & \cellcolor{bluegray} & \cellcolor{bluegray} & 245 \\ 
	\hline \hline 
	\end{tabularx}
	\caption{The number of independent constraints on the Yukawa coefficients from combining any two of the models, with the diagonal being the constraints from the individual models.}
	\label{tab:yukawa_constraints}
\end{table}

For each of the models used, we have computed 4-loop gauge and/or 3-loop Yukawa \befs in the regular \msbar scheme. 
The models were then implemented in \texttt{RGBeta} to evaluate the model-specific \bef parametrization of Eqs.~(\ref{eq:beta_gauge_l}--\ref{eq:beta_quartic_l}). 
The two expressions for the \befs were then identified on a term-by-term basis to extract linear constraints on the \bef coefficients needed to reproduce the full perturbative result.     

The models used to constrain the \bef coefficients are mostly SM-like and so many of the constraints are degenerate between them. 
To clarify how the individual model constraints supplement each other, Tab.~\ref{tab:gauge_constraints} shows the number of independent constraints on the gauge \bef obtained from combining any two of the models (the diagonal then being the number of constraints from just the one model). 
In total all the models contribute 198 constraints on the 202 coefficients for the gauge \bef, corresponding to all but the $ \gamma_5 $-odd terms.
Similarly, Tab.~\ref{tab:yukawa_constraints} shows the number of independent constraints on the Yukawa \bef coefficients. 
Between all the models, we obtain a total of 292 constraints on the 308 Yukawa coefficients. 

As discussed in Sec.~\ref{sec:WCCs}, the WCCs provide a crucial set of constraints, similar in number to those obtained from the Yukawa \befs. 
Furthermore, these are the only constraints linking gauge and Yukawa coefficients, and they completely predict the $ \gamma_5 $-odd gauge coefficients from the 3-loop Yukawa \bef. 
The breakdown of how the various sources of constraints supplement each other is detailed in Tab.~\ref{tab:combined_constraints}. 

With all available information, we manage to extract all 510 coefficients of the 4-loop gauge and 3-loop Yukawa \befs. 
It is worth pointing out that there is a large number of redundant constraints obtained from the various models as well as from the WCCs. The consistency of all these redundant
conditions serves as a strong crosscheck of our results. 
For future reference, we have included the results for all coefficients in the appendix.

We find full agreement with the coefficients determined in Ref.~\cite{Bednyakov:2021qxa}. However, while we use a similar methodology, our determination of the coefficients differs significantly
from Ref.~\cite{Bednyakov:2021qxa}. Our choice of models allows for a full determination of the 198 $\gamma_5$-even gauge \bef coefficients, without needing to make use of the WCCs. Additionally, we do not rely on the conjectured symmetry of $ T^{IJ} $, thus providing two more constraints and verifying this conjecture to 4--3--2-loop order.

\begin{table}
	\centering
	\begin{tabularx}{.55\textwidth}{| Y | Y Y Y|}
	\hline \hline 
	Combined & Gauge & Yukawa & WCCs \\ \hline 
	Gauge& 198 & 490 & 424 \\ 
	Yukawa & \cellcolor{bluegray} & 292 & 438 \\ 
	WCCs & \cellcolor{bluegray} & \cellcolor{bluegray} & 266\\ 
	\hline \hline 
	\end{tabularx}
	\caption{The number of independent constraints on the \bef coefficients obtained from combining various sets of constraints.}
	\label{tab:combined_constraints}
\end{table}

As discussed in Section~\ref{sec:WCCs}, the determination of the \befs completely fixes the coefficients of $ \upsilon $. 
This work, therefore, also determines the flavor-improved 3-loop Yukawa \bef, $ B^{(3)}_{aij} $. 
Furthermore, $ \upsilon^{(3)}_{ab} $ can be used to determine $ B^{(3)}_{abcd} $ if and when the 3-loop quartic \bef is calculated with the Hermitian renormalization conditions used here.

\section{Conclusion} \label{sec::conc}
We have finalized the computation of the \befs of the most general gauge-Yukawa theory at 4--3--2-loop order and find full agreement with the findings of Ref.~\cite{Bednyakov:2021qxa}. 
In contrast to Ref.~\cite{Bednyakov:2021qxa} however, we do not rely on the symmetry of $T_{IJ}$ and use a different set of models to fix the coefficients.
These latest results stand apart from the direct calculations of the general \befs at lower loop order~\cite{Machacek:1983fi,Machacek:1983tz,Machacek:1984zw,Jack:1984vj,Pickering:2001aq}; the 4-loop gauge and 3-loop Yukawa \befs were determined with a new methodology, filling in the coefficients of a general parametrization with several partial results along with the WCCs.

We have made the results of our computation including all TSs and coefficients directly available in the appendix. Appreciating, however, that it requires a lot of effort to utilize this information directly in computations, we have also implemented them directly in the latest update of \texttt{RGBeta} (v1.1.0)~\cite{Thomsen:2021ncy}. 

With the 4--3--2-loop \befs now fully determined, our next goal is the exploration of the \befs at 5--4--3-loop order, necessitating further developments of the WCCs and \bef parametrization.
The $\gamma_5$-odd sector at this order is of particular interest, as it is not clear if the \emph{miraculous} full determination of all $\gamma_5$-odd coefficients via the computation of unambiguous terms can be repeated.

\subsection*{Acknowledgments}
We would like to thank A. Bednyakov and A. Pikelner for correspondence regarding their recent work~\cite{Bednyakov:2021qxa}.
The work of AET has received funding from the Swiss National Science Foundation (SNF) through the Eccellenza Professorial Fellowship ``Flavor Physics at the High Energy Frontier'' project number 186866.
FH acknowledges support by the Alexander von Humboldt foundation. This document was prepared using the resources of the Fermi National Accelerator Laboratory (Fermilab), a U.S. Department of Energy, Office of Science,
HEP User Facility. Fermilab is managed by Fermi Research Alliance, LLC (FRA), acting under Contract No. DE-AC02-07CH11359.
The work of JD was supported by the Science and Technology Facilities Council (STFC) under the Consolidated Grant ST/T00102X/1.

\app[Appendix]
The appendix reports the explicit TS parametrization of the 4-loop gauge and 3-loop Yukawa \befs, as well as the 3-loop $ \upsilon $-function, determined diagrammatically in Ref.~\cite{Poole:2019kcm}. 
We provide the values of all the associated coefficients as determined in this work. 
For similar parametrization and coefficients of the 3--2--2 \befs, we refer the reader to Ref.~\cite{Poole:2019kcm}. 
Between them, Ref.~\cite{Poole:2019kcm} and this appendix provide all TSs and \msbar coefficients of the 4--3--2-loop \befs.

\subsection{Tensor notation}
We define some frequently occurring tensors. The fermion indices are always treated in matrix notation and never made explicit. Tildes on more advanced objects are always defined in the usual manner: $ \tilde{X} = \sigma_1 X \sigma_1 $ in the space of fermion indices. 
For compactness, gauge indices are always contracted with an implicit $ a_{AB} $, while one should rescale the structure constants according to $ F^{ABC} \to a^{\eminus 1}_{AD}F^{DBC} $.  
Several 1- and 2-loop 2-point functions occurring as substructures in the \befs are used throughout. They are: 

\noindent 1-loop gauge:\footnote{Be advised that $ S_2(F) $ is twice the trace normalization of the fermions, due to both fermions and their complex conjugates automatically being included in the notation.}
    \begin{align}
    [S_2(F)]_{AB} &= \Tr{T^{A} T^{B}}\,, & 
    [S_2(S)]_{AB} &= [T_\phi^{A} T_\phi^{B}]_{aa}\,, &
    [C_2(G)]_{AB} &= F^{ACD} F^{CDB}.    & 
    \end{align}
\noindent 1-loop fermion:
    \begin{align}
    C_2(F) &= T^{A}T^{A}, & Y_2(F) &= y_a \tilde{y}_a\,.
    \end{align}
1-loop scalar:
    \begin{align}
    [C_2(S)]_{ab} &= [T_\phi^{A} T_\phi^{A}]_{ab}\,, & [Y_2(S)]_{ab} &= \Tr{y_a \tilde{y}_b}\,.
    \end{align}
2-loop gauge:
    \begin{align}
    [S_2(F,C_F)]_{AB} &= \Tr{T^{A} T^{B} C_2(F)}\,, &
    [S_2(F,Y_F)]_{AB} &= \Tr{T^{A} T^{B} \widetilde{Y}_2(F)}\,, \nonumber \\
    [S_2(S,C_S)]_{AB} &= [T^{A}_\phi T^{B}_\phi C_2(S)]_{aa}\,, &
    [S_2(S,Y_S)]_{AB} &= [T^{A}_\phi T^{B}_\phi Y_2(S)]_{aa}\,.
    \end{align}
2-loop fermion:
    \begin{align}
    C_2(F,G) &= T^{A}T^{B} [C_2(G)]_{AB}\,, & 
    C_2(F,S) &= T^{A}T^{B} [S_2(S)]_{AB}\,,  \nonumber \\ 
    C_2(F,F) &= T^{A}T^{B} [S_2(F)]_{AB}\,, &
    \end{align}
2-loop scalar:
    \begin{align}
    [C_2(S,G)]_{ab} &= [T_\phi^{A} T_\phi^{B}]_{ab} [C_2(G)]_{AB}\,, &
    [C_2(S,S)]_{ab} &= [T_\phi^{A} T_\phi^{B}]_{ab} [S_2(S)]_{AB}\,, \nonumber \\
    [C_2(S,F)]_{ab} &= [T_\phi^{A} T_\phi^{B}]_{ab} [S_2(F)]_{AB}\,, &
    [\Lambda_2]_{ab} &= \lambda_{acde} \lambda_{bcde}\,.
    \end{align}
To keep the explicit indices to a minimum, we stick with the convention of using implicit matrix multiplication for fermion indices. 
For the scalars, we have at times used shorthand $ [ \cdots]_{ab} $ to denote a matrix product of tensors with 2 scalar indices contracted with open indices $ a,b $. 
Similarly, a scalar trace is denoted $ [ \cdots]_{aa} $.
Having defined the 2-point structures, we can proceed with the parametrization of the \befs.

\subsection{Gauge \bef}
The $ \gamma_5 $-even TSs of the 4-loop gauge \bef are  
{\small
\begin{align*}
\beta^{(4)}_{AB} =& \\
\cofg{4}{1} &F^{AGI} F^{CDE} F^{CFG} F^{DHI} F^{EJK} F^{LHJ} F^{LMB}  & \!\!\!\!\!F^{MFK} \,\,\,\,\,
+\cofg{4}{2} &\big[T_\Phi^C T_\Phi^B T_\Phi^E T_\Phi^D \big]_{bb} \big[T_\Phi^C T_\Phi^D T_\Phi^E T_\Phi^A \big]_{aa}  \\ 
+\cofg{4}{3} &F^{ACF} F^{CDE} \big[T_\Phi^D T_\Phi^B T_\Phi^G T_\Phi^F T_\Phi^E T_\Phi^G \big]_{aa}  & 
+\cofg{4}{4} &\Tr{T^C T^D T^E T^A } \big[T_\Phi^C T_\Phi^D T_\Phi^E T_\Phi^B \big]_{aa}  \\ 
+\cofg{4}{5} &\Tr{T^C T^D T^E T^B } \Tr{T^D T^C T^A T^E }  & 
+\cofg{4}{6} &\Tr{T^C T^D T^A T^C T^E T^B T^D T^E }  \\ 
+\cofg{4}{7} &[S_2(S,C_S)]_{CD} \big[T_\Phi^C T_\Phi^D T_\Phi^A T_\Phi^B \big]_{aa}  & 
+\cofg{4}{8} &[S_2(F,C_F)]_{CD} \big[T_\Phi^C T_\Phi^A T_\Phi^B T_\Phi^D \big]_{aa}  \\ 
+\cofg{4}{9} &[S_2(S,C_S)]_{CD} \Tr{T^D T^C T^A T^B }  & 
+\cofg{4}{10} &[S_2(F,C_F)]_{CD} \Tr{T^B T^D T^C T^A }  \\ 
+\cofg{4}{11} &\Tr{T^B C_2(F) C_2(F) C_2(F) T^A }  & 
+\cofg{4}{12} &\big[C_2(S) C_2(S) C_2(S) T_\Phi^A T_\Phi^B \big]_{aa}  \\ 
+\cofg{4}{13} &\Tr{T^B C_2(F) C_2(F,G) T^A }  & 
+\cofg{4}{14} &\big[C_2(S) C_2(S,G) T_\Phi^A T_\Phi^B \big]_{aa}  \\ 
+\cofg{4}{15} &\Tr{C_2(F,F) C_2(F) T^A T^B }  & 
+\cofg{4}{16} &[C_2(S,F)]_{ab} (T_\Phi^A)_{bc} \big[T_\Phi^B C_2(S) \big]_{ca}  \\ 
+\cofg{4}{17} &\Tr{C_2(F,S) C_2(F) T^A T^B }  & 
+\cofg{4}{18} &[C_2(S,S)]_{ab} \big[C_2(S) T_\Phi^B T_\Phi^A \big]_{ba}  \\ 
+\cofg{4}{19} &[C_2(G)]_{CB} \Tr{T^A C_2(F) C_2(F) T^C }  & 
+\cofg{4}{20} &[C_2(G)]_{CA} \big[C_2(S) C_2(S) T_\Phi^C T_\Phi^B \big]_{aa}  \\ 
+\cofg{4}{21} &[S_2(F)]_{CD} [S_2(F)]_{ED} \Tr{T^C T^E T^A T^B }  & 
+\cofg{4}{22} &[S_2(F)]_{CD} [S_2(S)]_{ED} \Tr{T^E T^C T^A T^B }  \\ 
+\cofg{4}{23} &[S_2(S)]_{CD} [S_2(S)]_{ED} \Tr{T^C T^E T^A T^B }  & 
+\cofg{4}{24} &[S_2(S)]_{CD} [S_2(S)]_{ED} \big[T_\Phi^C T_\Phi^A T_\Phi^B T_\Phi^E \big]_{aa}  \\ 
+\cofg{4}{25} &[C_2(G)]_{CD} [S_2(F)]_{CE} \Tr{T^E T^D T^A T^B }  & 
+\cofg{4}{26} &[C_2(G)]_{CD} [S_2(S)]_{CE} \big[T_\Phi^D T_\Phi^B T_\Phi^A T_\Phi^E \big]_{aa}  \\ 
+\cofg{4}{27} &[S_2(F)]_{CD} [S_2(S)]_{DE} \big[T_\Phi^C T_\Phi^B T_\Phi^A T_\Phi^E \big]_{aa}  & 
+\cofg{4}{28} &[C_2(G)]_{CD} [S_2(S)]_{CE} \Tr{T^E T^D T^A T^B }  \\ 
+\cofg{4}{29} &[C_2(G)]_{CD} [C_2(G)]_{CE} \Tr{T^B T^E T^D T^A }  & 
+\cofg{4}{30} &[C_2(G)]_{CD} [C_2(G)]_{CE} \big[T_\Phi^D T_\Phi^A T_\Phi^B T_\Phi^E \big]_{aa}  \\ 
+\cofg{4}{31} &[C_2(G)]_{CD} [S_2(F)]_{CE} \big[T_\Phi^D T_\Phi^E T_\Phi^A T_\Phi^B \big]_{aa}  & 
+\cofg{4}{32} &[S_2(F)]_{CD} [S_2(F)]_{CE} \big[T_\Phi^D T_\Phi^A T_\Phi^B T_\Phi^E \big]_{aa}  \\ 
+\cofg{4}{33} &[C_2(G)]_{CB} [C_2(G)]_{CD} [S_2(F,C_F)]_{DA}  & 
+\cofg{4}{34} &[C_2(G)]_{CA} [C_2(G)]_{CD} [S_2(S,C_S)]_{BD}  \\ 
+\cofg{4}{35} &[C_2(G)]_{CB} \Tr{T^A T^C C_2(F,G) }  & 
+\cofg{4}{36} &[C_2(G)]_{CA} \big[C_2(S,G) T_\Phi^B T_\Phi^C \big]_{aa}  \\ 
+\cofg{4}{37} &[C_2(G)]_{CB} \Tr{C_2(F,F) T^C T^A }  & 
+\cofg{4}{38} &[C_2(G)]_{CA} [C_2(S,F)]_{ab} \big[T_\Phi^B T_\Phi^C \big]_{ba}  \\ 
+\cofg{4}{39} &[C_2(G)]_{CB} \Tr{C_2(F,S) T^C T^A }  & 
+\cofg{4}{40} &[C_2(G)]_{CA} [C_2(S,S)]_{ab} \big[T_\Phi^C T_\Phi^B \big]_{ba}  \\ 
+\cofg{4}{41} &[C_2(G)]_{CB} [S_2(F)]_{DA} [S_2(F,C_F)]_{CD}  & 
+\cofg{4}{42} &[C_2(G)]_{CB} [S_2(F)]_{DA} [S_2(S,C_S)]_{DC}  \\ 
+\cofg{4}{43} &[C_2(G)]_{CA} [S_2(F,C_F)]_{CD} [S_2(S)]_{DB}  & 
+\cofg{4}{44} &[C_2(G)]_{CA} [S_2(S)]_{DB} [S_2(S,C_S)]_{DC}  \\ 
+\cofg{4}{45} &[C_2(G)]_{CB} [C_2(G)]_{CD} [C_2(G)]_{DE} [C_2(G)]_{EA}  & 
+\cofg{4}{46} &[C_2(G)]_{CD} [C_2(G)]_{CE} [C_2(G)]_{DB} [S_2(F)]_{EA}  \\ 
+\cofg{4}{47} &[C_2(G)]_{CD} [C_2(G)]_{CE} [C_2(G)]_{DA} [S_2(S)]_{EB}  & 
+\cofg{4}{48} &[C_2(G)]_{CB} [S_2(F)]_{CD} [S_2(F)]_{DE} [S_2(F)]_{EA}  \\ 
+\cofg{4}{49} &[C_2(G)]_{CB} [C_2(G)]_{CD} [S_2(F)]_{DE} [S_2(F)]_{EA}  & 
+\cofg{4}{50} &[C_2(G)]_{CA} [S_2(S)]_{CD} [S_2(S)]_{DE} [S_2(S)]_{EB}  \\ 
+\cofg{4}{51} &[C_2(G)]_{CA} [C_2(G)]_{CD} [S_2(S)]_{DE} [S_2(S)]_{EB}  & 
+\cofg{4}{52} &[C_2(G)]_{CA} [C_2(G)]_{CD} [S_2(F)]_{DE} [S_2(S)]_{EB}  \\ 
+\cofg{4}{53} &[C_2(G)]_{CD} [S_2(F)]_{CE} [S_2(F)]_{DA} [S_2(S)]_{EB}  & 
+\cofg{4}{54} &[C_2(G)]_{CD} [S_2(F)]_{CA} [S_2(S)]_{DE} [S_2(S)]_{EB}  \\ 
+\cofg{4}{55} &F^{ACF} F^{CDE} \lambda_{abcd} \big[T_\Phi^D T_\Phi^B \big]_{ad} \big[T_\Phi^E T_\Phi^F \big]_{bc}  & 
+\cofg{4}{56} &\lambda_{abcd} (T_\Phi^C)_{de} \big[T_\Phi^A C_2(S) \big]_{ea} \big[T_\Phi^C T_\Phi^B \big]_{bc}  \\ 
+\cofg{4}{57} &\lambda_{abcd} \big[T_\Phi^C T_\Phi^D \big]_{ac} \big[T_\Phi^C T_\Phi^D T_\Phi^A T_\Phi^B \big]_{db}  & 
+\cofg{4}{58} &[C_2(G)]_{CD} \lambda_{abcd} \big[T_\Phi^C T_\Phi^A \big]_{bc} \big[T_\Phi^D T_\Phi^B \big]_{ad}  \\ 
+\cofg{4}{59} &\lambda_{abcd} [S_2(S)]_{CD} \big[T_\Phi^A T_\Phi^C \big]_{ad} \big[T_\Phi^B T_\Phi^D \big]_{bc}  & 
+\cofg{4}{60} &[C_2(G)]_{CA} \lambda_{abcd} \big[T_\Phi^C T_\Phi^D \big]_{ad} \big[T_\Phi^D T_\Phi^B \big]_{bc}  \\ 
+\cofg{4}{61} &\lambda_{abcd} [S_2(F)]_{CD} \big[T_\Phi^C T_\Phi^A \big]_{bc} \big[T_\Phi^D T_\Phi^B \big]_{ad}  & 
+\cofg{4}{62} &\lambda_{abcd} \lambda_{cefd} \big[T_\Phi^A T_\Phi^C \big]_{eb} \big[T_\Phi^C T_\Phi^B \big]_{af}  \\ 
+\cofg{4}{63} &\lambda_{abcd} \lambda_{cefd} \big[T_\Phi^A T_\Phi^C \big]_{ef} \big[T_\Phi^C T_\Phi^B \big]_{ab}  & 
+\cofg{4}{64} &\big[T_\Phi^B T_\Phi^A C_2(S) \Lambda_2 \big]_{ab}  \\ 
+\cofg{4}{65} &[C_2(S)]_{ab} \lambda_{acde} \lambda_{bdfe} \big[T_\Phi^A T_\Phi^B \big]_{cf}  & 
+\cofg{4}{66} &[C_2(G)]_{CA} \big[T_\Phi^B \Lambda_2 T_\Phi^C \big]_{ab}  \\ 
+\cofg{4}{67} &\lambda_{abcd} \lambda_{aefc} \lambda_{bghd} (T_\Phi^A)_{ge} (T_\Phi^B)_{hf}  & 
+\cofg{4}{68} &\lambda_{abcd} \lambda_{aefc} \lambda_{bfgd} \big[T_\Phi^A T_\Phi^B \big]_{ge}  \\ 
+\cofg{4}{69} &\Tr{T^C T^D T^A T^C \tilde{y}_a \widetilde{T}^B \widetilde{T}^D y_a }  & 
+\cofg{4}{70} &[S_2(F,Y_F)]_{CD} \big[T_\Phi^C T_\Phi^A T_\Phi^B T_\Phi^D \big]_{aa}  \\ 
+\cofg{4}{71} &[S_2(S,Y_S)]_{CD} \big[T_\Phi^C T_\Phi^A T_\Phi^B T_\Phi^D \big]_{aa}  & 
+\cofg{4}{72} &[S_2(F,Y_F)]_{CD} \Tr{T^B T^D T^C T^A }  \\ 
+\cofg{4}{73} &[S_2(S,Y_S)]_{CD} \Tr{T^C T^D T^A T^B }  & 
+\cofg{4}{74} &\Tr{C_2(F) C_2(F) T^A T^B \widetilde{Y}_2(F) }  \\ 
+\cofg{4}{75} &\Tr{\tilde{y}_a \widetilde{C}_2(F) y_b C_2(F) } \big[T_\Phi^A T_\Phi^B \big]_{ab}  & 
+\cofg{4}{76} &\Tr{T^B C_2(F) \tilde{y}_a \widetilde{C}_2(F) y_a T^A }  \\ 
+\cofg{4}{77} &\Tr{\tilde{y}_a y_b T^A T^B } \big[C_2(S) C_2(S) \big]_{ab}  & 
+\cofg{4}{78} &\Tr{\widetilde{C}_2(F) \widetilde{C}_2(F) y_a T^A T^B \tilde{y}_a }  \\ 
+\cofg{4}{79} &\Tr{C_2(F) C_2(F) \tilde{y}_a y_b } \big[T_\Phi^A T_\Phi^B \big]_{ab}  & 
+\cofg{4}{80} &[C_2(S)]_{ab} \Tr{\tilde{y}_a \widetilde{C}_2(F) y_b T^A T^B }  \\ 
+\cofg{4}{81} &\Tr{y_a C_2(F) \tilde{y}_b } \big[C_2(S) T_\Phi^B T_\Phi^A \big]_{ab}  & 
+\cofg{4}{82} &[C_2(S)]_{ab} \Tr{T^B C_2(F) \tilde{y}_a y_b T^A }  \\ 
+\cofg{4}{83} &\big[T_\Phi^A T_\Phi^B C_2(S) C_2(S) Y_2(S) \big]_{ab}  & 
+\cofg{4}{84} &\Tr{T^B \widetilde{Y}_2(F) C_2(F,G) T^A }  \\ 
+\cofg{4}{85} &\Tr{C_2(F,F) \widetilde{Y}_2(F) T^A T^B }  & 
+\cofg{4}{86} &\Tr{C_2(F,S) \widetilde{Y}_2(F) T^A T^B }  \\ 
+\cofg{4}{87} &\Tr{\tilde{y}_a \widetilde{C}_2(F,G) y_a T^A T^B }  & 
+\cofg{4}{88} &[C_2(S,G)]_{ab} \Tr{\tilde{y}_a y_b T^A T^B }  \\ 
+\cofg{4}{89} &\Tr{y_a C_2(F,G) \tilde{y}_b } \big[T_\Phi^A T_\Phi^B \big]_{ab}  & 
+\cofg{4}{90} &\Tr{\tilde{y}_a \widetilde{C}_2(F,F) y_a T^A T^B }  \\ 
+\cofg{4}{91} &[C_2(S,F)]_{ab} \Tr{\tilde{y}_a y_b T^A T^B }  & 
+\cofg{4}{92} &\Tr{y_a C_2(F,F) \tilde{y}_b } \big[T_\Phi^A T_\Phi^B \big]_{ab}  \\ 
+\cofg{4}{93} &\Tr{\tilde{y}_a \widetilde{C}_2(F,S) y_a T^A T^B }  & 
+\cofg{4}{94} &[C_2(S,S)]_{ab} \Tr{\tilde{y}_a y_b T^A T^B }  \\ 
+\cofg{4}{95} &\Tr{y_a C_2(F,S) \tilde{y}_b } \big[T_\Phi^A T_\Phi^B \big]_{ab}  & 
+\cofg{4}{96} &[C_2(G)]_{CB} \Tr{T^A C_2(F) \widetilde{Y}_2(F) T^C }  \\ 
+\cofg{4}{97} &[C_2(G)]_{CB} \Tr{\tilde{y}_a \widetilde{C}_2(F) y_a T^A T^C }  & 
+\cofg{4}{98} &[C_2(G)]_{CA} \Tr{y_a C_2(F) \tilde{y}_b } \big[T_\Phi^C T_\Phi^B \big]_{ab}  \\ 
+\cofg{4}{99} &[C_2(G)]_{CB} [C_2(S)]_{ab} \Tr{y_a T^C T^A \tilde{y}_b }  & 
+\cofg{4}{100} &\big[T_\Phi^A T_\Phi^B C_2(S,G) Y_2(S) \big]_{ab}  \\ 
+\cofg{4}{101} &[C_2(S,F)]_{ab} \big[Y_2(S) T_\Phi^A T_\Phi^B \big]_{ca}  & 
+\cofg{4}{102} &[C_2(S,S)]_{ab} \big[T_\Phi^B T_\Phi^A Y_2(S) \big]_{bc}  \\ 
+\cofg{4}{103} &[C_2(G)]_{CA} \big[T_\Phi^B T_\Phi^C C_2(S) Y_2(S) \big]_{ab}  & 
+\cofg{4}{104} &[C_2(G)]_{CB} [C_2(G)]_{CD} [S_2(F,Y_F)]_{DA}  \\ 
+\cofg{4}{105} &[C_2(G)]_{CB} [S_2(F)]_{DA} [S_2(F,Y_F)]_{CD}  & 
+\cofg{4}{106} &[C_2(G)]_{CA} [S_2(F,Y_F)]_{CD} [S_2(S)]_{DB}  \\ 
+\cofg{4}{107} &[C_2(G)]_{CA} [C_2(G)]_{CD} [S_2(S,Y_S)]_{DB}  & 
+\cofg{4}{108} &[C_2(G)]_{CB} [S_2(F)]_{DA} [S_2(S,Y_S)]_{CD}  \\ 
+\cofg{4}{109} &[C_2(G)]_{CA} [S_2(S)]_{DB} [S_2(S,Y_S)]_{CD}  & 
+\cofg{4}{110} &\lambda_{abcd} \Tr{y_c T^C T^A \tilde{y}_d } \big[T_\Phi^C T_\Phi^B \big]_{ab}  \\ 
+\cofg{4}{111} &\lambda_{abcd} \big[T_\Phi^C T_\Phi^B \big]_{bc} \big[T_\Phi^C T_\Phi^A Y_2(S) \big]_{da}  & 
+\cofg{4}{112} &[\Lambda_2]_{ab} \Tr{\tilde{y}_a y_b T^A T^B }  \\ 
+\cofg{4}{113} &\big[\Lambda_2 T_\Phi^B T_\Phi^A Y_2(S) \big]_{aa}  & 
+\cofg{4}{114} &\lambda_{abcd} \lambda_{cefd} [Y_2(S)]_{ae} \big[T_\Phi^A T_\Phi^B \big]_{bf}  \\ 
+\cofg{4}{115} &\Tr{y_a T^C T^A \tilde{y}_a y_b T^B T^C \tilde{y}_b }  & 
+\cofg{4}{116} &\Tr{y_a T^C T^A \tilde{y}_b } \Tr{y_a T^C T^B \tilde{y}_b }  \\ 
+\cofg{4}{117} &\Tr{\tilde{y}_a \widetilde{T}^C \widetilde{T}^A y_b T^C \tilde{y}_a y_b T^B }  & 
+\cofg{4}{118} &[C_2(S)]_{ab} \Tr{\tilde{y}_a y_c T^A \tilde{y}_c y_b T^B }  \\ 
+\cofg{4}{119} &\Tr{y_a T^C \tilde{y}_b y_c T^C \tilde{y}_c } \big[T_\Phi^A T_\Phi^B \big]_{ab}  & 
+\cofg{4}{120} &\Tr{\tilde{y}_a \widetilde{C}_2(F) y_b T^A \tilde{y}_b y_a T^B }  \\ 
+\cofg{4}{121} &\Tr{T^B C_2(F) \tilde{y}_a y_b T^A \tilde{y}_b y_a }  & 
+\cofg{4}{122} &\Tr{\tilde{y}_a \widetilde{T}^C y_a T^A T^B \tilde{y}_b \widetilde{T}^C y_b }  \\ 
+\cofg{4}{123} &\Tr{y_a T^C T^A T^B \tilde{y}_a y_b T^C \tilde{y}_b }  & 
+\cofg{4}{124} &[C_2(S)]_{ab} \Tr{\tilde{y}_a y_c \tilde{y}_b y_d } \big[T_\Phi^A T_\Phi^B \big]_{dc}  \\ 
+\cofg{4}{125} &\Tr{\tilde{y}_a y_b \tilde{y}_c y_b } \big[C_2(S) T_\Phi^B T_\Phi^A \big]_{ac}  & 
+\cofg{4}{126} &\Tr{y_a C_2(F) \tilde{y}_b y_c \tilde{y}_b } \big[T_\Phi^A T_\Phi^B \big]_{ca}  \\ 
+\cofg{4}{127} &[C_2(S)]_{ab} \Tr{\tilde{y}_a y_c T^A T^B \tilde{y}_b y_c }  & 
+\cofg{4}{128} &\Tr{y_a C_2(F) \tilde{y}_b y_a T^A T^B \tilde{y}_b }  \\ 
+\cofg{4}{129} &\Tr{\tilde{y}_a \widetilde{C}_2(F) y_b T^A T^B \tilde{y}_a y_b }  & 
+\cofg{4}{130} &\Tr{T^B C_2(F) \tilde{y}_a y_b \tilde{y}_a y_b T^A }  \\ 
+\cofg{4}{131} &\Tr{y_a C_2(F) \tilde{y}_b } \Tr{\tilde{y}_b y_a T^A T^B }  & 
+\cofg{4}{132} &\Tr{\tilde{y}_a y_b T^A T^B } \big[C_2(S) Y_2(S) \big]_{ac}  \\ 
+\cofg{4}{133} &\Tr{y_a C_2(F) \tilde{y}_b } \big[T_\Phi^A T_\Phi^B Y_2(S) \big]_{ac}  & 
+\cofg{4}{134} &\Tr{\tilde{y}_a \widetilde{C}_2(F) y_b T^A T^B } [Y_2(S)]_{ab}  \\ 
+\cofg{4}{135} &\Tr{T^B C_2(F) \tilde{y}_a y_b T^A } [Y_2(S)]_{ab}  & 
+\cofg{4}{136} &\Tr{\tilde{y}_a Y_2(F) y_b } \big[C_2(S) T_\Phi^B T_\Phi^A \big]_{ab}  \\ 
+\cofg{4}{137} &[C_2(S)]_{ab} \Tr{\tilde{y}_a y_c \tilde{y}_d y_b } \big[T_\Phi^A T_\Phi^B \big]_{dc}  & 
+\cofg{4}{138} &\Tr{\widetilde{Y}_2(F) C_2(F) \tilde{y}_a y_b } \big[T_\Phi^A T_\Phi^B \big]_{ba}  \\ 
+\cofg{4}{139} &\Tr{y_a C_2(F) \tilde{y}_b Y_2(F) } \big[T_\Phi^A T_\Phi^B \big]_{ab}  & 
+\cofg{4}{140} &[C_2(S)]_{ab} \Tr{\tilde{y}_a Y_2(F) y_b T^A T^B }  \\ 
+\cofg{4}{141} &[C_2(S)]_{ab} \Tr{T^B \widetilde{Y}_2(F) \tilde{y}_a y_b T^A }  & 
+\cofg{4}{142} &\Tr{y_a C_2(F) \tilde{y}_a y_b \tilde{y}_c } \big[T_\Phi^A T_\Phi^B \big]_{bc}  \\ 
+\cofg{4}{143} &[C_2(S)]_{ab} \Tr{\tilde{y}_a y_c T^A T^B \tilde{y}_c y_b }  & 
+\cofg{4}{144} &\Tr{T^B \widetilde{Y}_2(F) C_2(F) \widetilde{Y}_2(F) T^A }  \\ 
+\cofg{4}{145} &\Tr{\widetilde{T}^B Y_2(F) y_a C_2(F) \tilde{y}_a \widetilde{T}^A }  & 
+\cofg{4}{146} &\Tr{Y_2(F) \widetilde{C}_2(F) y_a T^A T^B \tilde{y}_a }  \\ 
+\cofg{4}{147} &\Tr{T^B C_2(F) \tilde{y}_a Y_2(F) y_a T^A }  & 
+\cofg{4}{148} &\Tr{y_a C_2(F) \tilde{y}_a y_b T^A T^B \tilde{y}_b }  \\ 
+\cofg{4}{149} &(T_\Phi^B)_{ab} \big[T_\Phi^A Y_2(S) C_2(S) Y_2(S) \big]_{bc}  & 
+\cofg{4}{150} &[C_2(G)]_{CB} \Tr{y_a T^C \tilde{y}_a y_b T^A \tilde{y}_b }  \\ 
+\cofg{4}{151} &[C_2(G)]_{CA} \Tr{\tilde{y}_a y_b \tilde{y}_c y_b } \big[T_\Phi^C T_\Phi^B \big]_{ac}  & 
+\cofg{4}{152} &[C_2(G)]_{CB} \Tr{y_a T^C T^A \tilde{y}_b y_a \tilde{y}_b }  \\ 
+\cofg{4}{153} &[C_2(G)]_{CB} \Tr{y_a T^C T^A \tilde{y}_b } [Y_2(S)]_{ab}  & 
+\cofg{4}{154} &[C_2(G)]_{CA} \Tr{\tilde{y}_a Y_2(F) y_b } \big[T_\Phi^C T_\Phi^B \big]_{ab}  \\ 
+\cofg{4}{155} &[C_2(G)]_{CB} \Tr{\widetilde{T}^A Y_2(F) Y_2(F) \widetilde{T}^C }  & 
+\cofg{4}{156} &[C_2(G)]_{CB} \Tr{\tilde{y}_a Y_2(F) y_a T^A T^C }  \\ 
+\cofg{4}{157} &[C_2(G)]_{CA} \big[Y_2(S) T_\Phi^B T_\Phi^C Y_2(S) \big]_{ab}  & 
+\cofg{4}{158} &\lambda_{abcd} \Tr{\tilde{y}_d y_a T^A \tilde{y}_b y_c T^B }  \\ 
+\cofg{4}{159} &\lambda_{abcd} \Tr{\tilde{y}_d y_e \tilde{y}_c y_b } \big[T_\Phi^A T_\Phi^B \big]_{ea}  & 
+\cofg{4}{160} &\lambda_{abcd} \Tr{\tilde{y}_d y_a T^A T^B \tilde{y}_b y_c }  \\ 
+\cofg{4}{161} &\Tr{\tilde{y}_a y_b T^A \tilde{y}_c y_a \tilde{y}_b \widetilde{T}^B y_c }  & 
+\cofg{4}{162} &\Tr{\tilde{y}_a y_b \tilde{y}_c y_d \tilde{y}_b y_c } \big[T_\Phi^A T_\Phi^B \big]_{ad}  \\ 
+\cofg{4}{163} &\Tr{\tilde{y}_a y_b T^A T^B \tilde{y}_c y_a \tilde{y}_b y_c }  & 
+\cofg{4}{164} &\Tr{\tilde{y}_a y_b T^A \tilde{y}_a y_c T^B \tilde{y}_c y_b }  \\ 
+\cofg{4}{165} &\Tr{\tilde{y}_a y_b T^A \tilde{y}_c y_b \tilde{y}_a y_c T^B }  & 
+\cofg{4}{166} &\Tr{\tilde{y}_a y_b T^A \tilde{y}_b y_a \tilde{y}_c \widetilde{T}^B y_c }  \\ 
+\cofg{4}{167} &\Tr{\tilde{y}_a Y_2(F) y_b T^A \tilde{y}_b y_a T^B }  & 
+\cofg{4}{168} &\Tr{T^B \widetilde{Y}_2(F) \tilde{y}_a y_b T^A \tilde{y}_b y_a }  \\ 
+\cofg{4}{169} &\Tr{\tilde{y}_a y_b T^A \tilde{y}_b y_c T^B } [Y_2(S)]_{ac}  & 
+\cofg{4}{170} &\Tr{\tilde{y}_a y_b \tilde{y}_c y_d \tilde{y}_b y_d } \big[T_\Phi^A T_\Phi^B \big]_{ac}  \\ 
+\cofg{4}{171} &\Tr{\tilde{y}_a y_b \tilde{y}_c y_d \tilde{y}_c y_b } \big[T_\Phi^A T_\Phi^B \big]_{ad}  & 
+\cofg{4}{172} &\Tr{\tilde{y}_a y_b T^A T^B \tilde{y}_a y_c \tilde{y}_b y_c }  \\ 
+\cofg{4}{173} &\Tr{\tilde{y}_a y_b T^A T^B \tilde{y}_c y_a \tilde{y}_c y_b }  & 
+\cofg{4}{174} &\Tr{\tilde{y}_a y_b \tilde{y}_c y_b } \Tr{\tilde{y}_c y_a T^A T^B }  \\ 
+\cofg{4}{175} &\Tr{\tilde{y}_a Y_2(F) y_b \tilde{y}_c y_b } \big[T_\Phi^A T_\Phi^B \big]_{ac}  & 
+\cofg{4}{176} &\Tr{\tilde{y}_a y_b \tilde{y}_c y_b \tilde{y}_c y_d } \big[T_\Phi^A T_\Phi^B \big]_{ad}  \\ 
+\cofg{4}{177} &\Tr{y_a \widetilde{Y}_2(F) \tilde{y}_b y_a T^A T^B \tilde{y}_b }  & 
+\cofg{4}{178} &\Tr{\tilde{y}_a Y_2(F) y_b T^A T^B \tilde{y}_a y_b }  \\ 
+\cofg{4}{179} &\Tr{T^B \widetilde{Y}_2(F) \tilde{y}_a y_b \tilde{y}_a y_b T^A }  & 
+\cofg{4}{180} &\Tr{\tilde{y}_a y_b T^A T^B \tilde{y}_b y_c \tilde{y}_a y_c }  \\ 
+\cofg{4}{181} &\Tr{\tilde{y}_a y_b \tilde{y}_c y_b } \big[T_\Phi^A T_\Phi^B Y_2(S) \big]_{ad}  & 
+\cofg{4}{182} &\Tr{\tilde{y}_a y_b \tilde{y}_c y_d } [Y_2(S)]_{db} \big[T_\Phi^A T_\Phi^B \big]_{ac}  \\ 
+\cofg{4}{183} &\Tr{\tilde{y}_a y_b T^A T^B \tilde{y}_c y_b } [Y_2(S)]_{ac}  & 
+\cofg{4}{184} &\Tr{\tilde{y}_a Y_2(F) y_b } \Tr{\tilde{y}_b y_a T^A T^B }  \\ 
+\cofg{4}{185} &\Tr{\tilde{y}_a Y_2(F) y_b \widetilde{Y}_2(F) } \big[T_\Phi^A T_\Phi^B \big]_{ab}  & 
+\cofg{4}{186} &\Tr{\tilde{y}_a Y_2(F) y_a \tilde{y}_b y_c } \big[T_\Phi^A T_\Phi^B \big]_{bc}  \\ 
+\cofg{4}{187} &\Tr{T^B \widetilde{Y}_2(F) \tilde{y}_a Y_2(F) y_a T^A }  & 
+\cofg{4}{188} &\Tr{y_a \widetilde{Y}_2(F) \tilde{y}_a y_b T^A T^B \tilde{y}_b }  \\ 
+\cofg{4}{189} &\Tr{\tilde{y}_a Y_2(F) y_b } \big[T_\Phi^A T_\Phi^B Y_2(S) \big]_{ac}  & 
+\cofg{4}{190} &\Tr{\tilde{y}_a y_b \tilde{y}_c y_d } [Y_2(S)]_{cb} \big[T_\Phi^A T_\Phi^B \big]_{ad}  \\ 
+\cofg{4}{191} &\Tr{\tilde{y}_a Y_2(F) y_b T^A T^B } [Y_2(S)]_{ab}  & 
+\cofg{4}{192} &\Tr{T^B \widetilde{Y}_2(F) \tilde{y}_a y_b T^A } [Y_2(S)]_{ab}  \\ 
+\cofg{4}{193} &\Tr{\tilde{y}_a y_b T^A T^B \tilde{y}_b y_c } [Y_2(S)]_{ac}  & 
+\cofg{4}{194} &\Tr{\tilde{y}_a y_b T^A T^B } \big[Y_2(S) Y_2(S) \big]_{ac}  \\ 
+\cofg{4}{195} &\Tr{\tilde{y}_a Y_2(F) Y_2(F) y_b } \big[T_\Phi^A T_\Phi^B \big]_{ab}  & 
+\cofg{4}{196} &\Tr{\widetilde{T}^B Y_2(F) Y_2(F) Y_2(F) \widetilde{T}^A }  \\ 
+\cofg{4}{197} &\Tr{\tilde{y}_a Y_2(F) Y_2(F) y_a T^A T^B }  & 
+\cofg{4}{198} &[Y_2(S)]_{ab} \big[Y_2(S) T_\Phi^B T_\Phi^A Y_2(S) \big]_{ac} .
\end{align*}
{\normalsize There are an additional 4 $ \gamma_5 $-odd TSs:}
\begin{align*}
+\cofg{4}{199} &\Tr{T^A \tilde{y}_a y_b T^C \sigma_3} \Tr{\sigma_3 \widetilde{T}^C y_b \tilde{y}_a \widetilde{T}^B} &
+\cofg{4}{200} & \Tr{T^A \tilde{y}_a \widetilde{T}^C y_b  \sigma_3} \Tr{\sigma_3 y_b T^C \tilde{y}_a \widetilde{T}^B} \\
+\cofg{4}{201} & \Tr{T^A \tilde{y}_a y_b T^C \sigma_3} \Tr{\sigma_3 y_b \tilde{y}_a \widetilde{T}^C \widetilde{T}^B} & 
+\cofg{4}{202} & \Tr{T^A \tilde{y}_a y_b T^C \sigma_3} \Tr{\sigma_3 y_b T^C \tilde{y}_a \widetilde{T}^B} \,.
\end{align*}
}

The coefficients of the 4-loop gauge \bef are 
{\small
\begin{align*}
\cofg{4}{1} & = \dfrac{160}{9} - \dfrac{1408}{3} \zeta_3 & 
\cofg{4}{2} & = \dfrac{32}{3} \zeta_3 - \dfrac{14}{9} & 
\cofg{4}{3} & = \dfrac{640}{3} \zeta_3 - \dfrac{457}{9} & 
\cofg{4}{4} & = \dfrac{32}{3} \zeta_3 - \dfrac{128}{9} \\ 
\cofg{4}{5} & = \dfrac{88}{9} - \dfrac{64}{3} \zeta_3 & 
\cofg{4}{6} & = \dfrac{832}{3} \zeta_3 - \dfrac{256}{9} & 
\cofg{4}{7} & = 8 \zeta_3 - \dfrac{71}{3} & 
\cofg{4}{8} & = 16 \zeta_3 - \dfrac{95}{6} \\ 
\cofg{4}{9} & = 4 \zeta_3 - \dfrac{34}{3} & 
\cofg{4}{10} & = 8 \zeta_3 - \dfrac{23}{3} & 
\cofg{4}{11} & = \dfrac{49}{9} - \dfrac{832}{3} \zeta_3 & 
\cofg{4}{12} & = -\dfrac{3}{2} \\ 
\cofg{4}{13} & = \dfrac{92}{27} + \dfrac{2320}{9} \zeta_3 & 
\cofg{4}{14} & = \dfrac{13733}{108} + \dfrac{352}{9} \zeta_3 & 
\cofg{4}{15} & = \dfrac{38}{27} + \dfrac{16}{9} \zeta_3 & 
\cofg{4}{16} & = -\dfrac{181}{27} - \dfrac{32}{9} \zeta_3 \\ 
\cofg{4}{17} & = \dfrac{28}{27} + \dfrac{8}{9} \zeta_3 & 
\cofg{4}{18} & = -\dfrac{221}{27} - \dfrac{16}{9} \zeta_3 & 
\cofg{4}{19} & = \dfrac{158}{9} + \dfrac{832}{3} \zeta_3 & 
\cofg{4}{20} & = \dfrac{291}{4} \\ 
\cofg{4}{21} & = -\dfrac{77}{486} & 
\cofg{4}{22} & = -\dfrac{47}{486} & 
\cofg{4}{23} & = \dfrac{91}{1944} & 
\cofg{4}{24} & = \dfrac{323}{1944} \\ 
\cofg{4}{25} & = -\dfrac{769}{486} - \dfrac{88}{9} \zeta_3 & 
\cofg{4}{26} & = -\dfrac{15025}{1944} - \dfrac{20}{3} \zeta_3 & 
\cofg{4}{27} & = \dfrac{29}{486} & 
\cofg{4}{28} & = -\dfrac{539}{486} - \dfrac{44}{9} \zeta_3 \\ 
\cofg{4}{29} & = \dfrac{9271}{486} - \dfrac{448}{9} \zeta_3 & 
\cofg{4}{30} & = \dfrac{60659}{486} - \dfrac{88}{3} \zeta_3 & 
\cofg{4}{31} & = -\dfrac{12025}{972} - \dfrac{40}{3} \zeta_3 & 
\cofg{4}{32} & = -\dfrac{49}{486} \\ 
\cofg{4}{33} & = -\dfrac{463}{27} - \dfrac{1676}{9} \zeta_3 & 
\cofg{4}{34} & = \dfrac{5633}{108} + \dfrac{424}{9} \zeta_3 & 
\cofg{4}{35} & = \dfrac{707}{27} - \dfrac{1292}{9} \zeta_3 & 
\cofg{4}{36} & = \dfrac{4801}{54} - \dfrac{176}{9} \zeta_3 \\ 
\cofg{4}{37} & = \dfrac{4}{9} \zeta_3 - \dfrac{85}{27} & 
\cofg{4}{38} & = \dfrac{16}{9} \zeta_3 - \dfrac{208}{27} & 
\cofg{4}{39} & = \dfrac{2}{9} \zeta_3 - \dfrac{121}{54} & 
\cofg{4}{40} & = \dfrac{8}{9} \zeta_3 - \dfrac{641}{108} \\ 
\cofg{4}{41} & = \dfrac{28}{9} \zeta_3 - \dfrac{221}{54} & 
\cofg{4}{42} & = \dfrac{28}{9} \zeta_3 - \dfrac{1081}{108} & 
\cofg{4}{43} & = \dfrac{211}{108} - \dfrac{22}{9} \zeta_3 & 
\cofg{4}{44} & = \dfrac{53}{216} - \dfrac{4}{9} \zeta_3 \\ 
\cofg{4}{45} & = \dfrac{440}{9} \zeta_3 - \dfrac{151013}{243} & 
\cofg{4}{46} & = \dfrac{37223}{162} + \dfrac{836}{9} \zeta_3 & 
\cofg{4}{47} & = \dfrac{20873}{432} - \dfrac{136}{9} \zeta_3 & 
\cofg{4}{48} & = -\dfrac{53}{972} \\ 
\cofg{4}{49} & = -\dfrac{4229}{324} - \dfrac{4}{3} \zeta_3 & 
\cofg{4}{50} & = -\dfrac{59}{1944} & 
\cofg{4}{51} & = \dfrac{251}{1296} - \dfrac{10}{9} \zeta_3 & 
\cofg{4}{52} & = -\dfrac{661}{108} - \dfrac{26}{9} \zeta_3 \\ 
\cofg{4}{53} & = -\dfrac{7}{216} & 
\cofg{4}{54} & = -\dfrac{4}{81} & 
\cofg{4}{55} & = -\dfrac{27}{2} & 
\cofg{4}{56} & = 17 \\ 
\cofg{4}{57} & = \dfrac{3}{2} & 
\cofg{4}{58} & = -\dfrac{37}{36} & 
\cofg{4}{59} & = -\dfrac{5}{36} & 
\cofg{4}{60} & = \dfrac{1}{2} \\ 
\cofg{4}{61} & = \dfrac{1}{18} & 
\cofg{4}{62} & = -\dfrac{5}{3} & 
\cofg{4}{63} & = -\dfrac{1}{12} & 
\cofg{4}{64} & = -\dfrac{11}{24} \\ 
\cofg{4}{65} & = -\dfrac{13}{8} & 
\cofg{4}{66} & = -\dfrac{1}{9} & 
\cofg{4}{67} & = -\dfrac{1}{12} & 
\cofg{4}{68} & = \dfrac{1}{8} \\ 
\cofg{4}{69} & = -36 & 
\cofg{4}{70} & = \dfrac{17}{4} & 
\cofg{4}{71} & = \dfrac{3}{2} & 
\cofg{4}{72} & = 2 \\ 
\cofg{4}{73} & = \dfrac{3}{4} & 
\cofg{4}{74} & = 25 \zeta_3 - \dfrac{59}{24} & 
\cofg{4}{75} & = -\dfrac{33}{2} - 6 \zeta_3 & 
\cofg{4}{76} & = \dfrac{49}{4} + 6 \zeta_3 \\ 
\cofg{4}{77} & = 15 \zeta_3 - 28 & 
\cofg{4}{78} & = \dfrac{593}{24} + 5 \zeta_3 & 
\cofg{4}{79} & = 1 - 6 \zeta_3 & 
\cofg{4}{80} & = -36 \zeta_3 \\ 
\cofg{4}{81} & = 10 \zeta_3 - \dfrac{55}{3} & 
\cofg{4}{82} & = 29 - 36 \zeta_3 & 
\cofg{4}{83} & = -\dfrac{27}{2} & 
\cofg{4}{84} & = -\dfrac{3445}{216} - \dfrac{7}{9} \zeta_3 \\ 
\cofg{4}{85} & = \dfrac{71}{216} + \dfrac{8}{9} \zeta_3 & 
\cofg{4}{86} & = \dfrac{359}{432} + \dfrac{4}{9} \zeta_3 & 
\cofg{4}{87} & = \dfrac{169}{9} \zeta_3 - \dfrac{5093}{216} & 
\cofg{4}{88} & = 3 \zeta_3 - \dfrac{1259}{36} \\ 
\cofg{4}{89} & = \dfrac{68}{9} - \dfrac{22}{3} \zeta_3 & 
\cofg{4}{90} & = \dfrac{463}{216} - \dfrac{8}{9} \zeta_3 & 
\cofg{4}{91} & = \dfrac{97}{36} & 
\cofg{4}{92} & = \dfrac{2}{3} \zeta_3 - \dfrac{4}{9} \\ 
\cofg{4}{93} & = \dfrac{679}{432} - \dfrac{4}{9} \zeta_3 & 
\cofg{4}{94} & = \dfrac{43}{18} & 
\cofg{4}{95} & = \dfrac{1}{3} \zeta_3 - \dfrac{1}{18} & 
\cofg{4}{96} & = -\dfrac{37}{2} \\ 
\cofg{4}{97} & = \dfrac{15}{2} & 
\cofg{4}{98} & = 2 \zeta_3 - \dfrac{11}{3} & 
\cofg{4}{99} & = -41 & 
\cofg{4}{100} & = \dfrac{55}{9} \zeta_3 - \dfrac{520}{27} \\ 
\cofg{4}{101} & = \dfrac{77}{108} - \dfrac{5}{9} \zeta_3 & 
\cofg{4}{102} & = \dfrac{203}{216} - \dfrac{5}{18} \zeta_3 & 
\cofg{4}{103} & = -\dfrac{19}{2} & 
\cofg{4}{104} & = -\dfrac{823}{18} \\ 
\cofg{4}{105} & = \dfrac{109}{36} & 
\cofg{4}{106} & = \dfrac{43}{72} & 
\cofg{4}{107} & = \dfrac{4721}{216} + \dfrac{11}{9} \zeta_3 & 
\cofg{4}{108} & = -\dfrac{61}{216} - \dfrac{1}{9} \zeta_3 \\ 
\cofg{4}{109} & = -\dfrac{223}{432} - \dfrac{1}{18} \zeta_3 & 
\cofg{4}{110} & = -4 & 
\cofg{4}{111} & = -\dfrac{13}{6} & 
\cofg{4}{112} & = \dfrac{1}{4} \\ 
\cofg{4}{113} & = -\dfrac{1}{144} & 
\cofg{4}{114} & = \dfrac{11}{48} & 
\cofg{4}{115} & = \dfrac{20}{3} & 
\cofg{4}{116} & = 6 - 12 \zeta_3 \\ 
\cofg{4}{117} & = \dfrac{7}{3} - 24 \zeta_3 & 
\cofg{4}{118} & = \dfrac{38}{3} & 
\cofg{4}{119} & = \dfrac{2}{3} & 
\cofg{4}{120} & = -1 \\ 
\cofg{4}{121} & = -\dfrac{11}{3} & 
\cofg{4}{122} & = -\dfrac{1}{2} & 
\cofg{4}{123} & = -\dfrac{19}{6} & 
\cofg{4}{124} & = -\dfrac{31}{12} - 6 \zeta_3 \\ 
\cofg{4}{125} & = \dfrac{61}{9} - \dfrac{52}{3} \zeta_3 & 
\cofg{4}{126} & = 24 \zeta_3 - \dfrac{9}{2} & 
\cofg{4}{127} & = 2 + 36 \zeta_3 & 
\cofg{4}{128} & = \dfrac{131}{12} - 14 \zeta_3 \\ 
\cofg{4}{129} & = -\dfrac{539}{36} - \dfrac{16}{3} \zeta_3 & 
\cofg{4}{130} & = \dfrac{89}{9} - \dfrac{38}{3} \zeta_3 & 
\cofg{4}{131} & = \dfrac{87}{8} - 6 \zeta_3 & 
\cofg{4}{132} & = \dfrac{43}{12} + 6 \zeta_3 \\ 
\cofg{4}{133} & = \zeta_3 - \dfrac{1}{2} & 
\cofg{4}{134} & = -\dfrac{301}{72} - \dfrac{1}{3} \zeta_3 & 
\cofg{4}{135} & = \dfrac{1}{3} \zeta_3 - \dfrac{113}{72} & 
\cofg{4}{136} & = \dfrac{80}{9} - \dfrac{5}{3} \zeta_3 \\ 
\cofg{4}{137} & = \dfrac{11}{6} & 
\cofg{4}{138} & = \dfrac{3}{2} + \zeta_3 & 
\cofg{4}{139} & = \dfrac{1}{6} + \zeta_3 & 
\cofg{4}{140} & = 3 \zeta_3 - \dfrac{9}{4} \\ 
\cofg{4}{141} & = \dfrac{31}{12} & 
\cofg{4}{142} & = \dfrac{7}{6} & 
\cofg{4}{143} & = \dfrac{23}{3} - 3 \zeta_3 & 
\cofg{4}{144} & = \dfrac{113}{144} + \dfrac{4}{3} \zeta_3 \\ 
\cofg{4}{145} & = -\dfrac{131}{144} - \dfrac{4}{3} \zeta_3 & 
\cofg{4}{146} & = \dfrac{14}{3} \zeta_3 - \dfrac{779}{144} & 
\cofg{4}{147} & = \dfrac{229}{72} - \dfrac{5}{3} \zeta_3 & 
\cofg{4}{148} & = \dfrac{145}{48} - 3 \zeta_3 \\ 
\cofg{4}{149} & = \dfrac{203}{72} + \dfrac{1}{6} \zeta_3 & 
\cofg{4}{150} & = \dfrac{59}{6} & 
\cofg{4}{151} & = \dfrac{4}{3} \zeta_3 - \dfrac{115}{9} & 
\cofg{4}{152} & = \dfrac{97}{6} \\ 
\cofg{4}{153} & = \dfrac{25}{4} & 
\cofg{4}{154} & = \dfrac{10}{9} - \dfrac{1}{3} \zeta_3 & 
\cofg{4}{155} & = \dfrac{3}{4} & 
\cofg{4}{156} & = \dfrac{17}{12} \\ 
\cofg{4}{157} & = \dfrac{1}{3} \zeta_3 - \dfrac{373}{144} & 
\cofg{4}{158} & = 0 & 
\cofg{4}{159} & = \dfrac{1}{6} & 
\cofg{4}{160} & = -\dfrac{5}{2} \\ 
\cofg{4}{161} & = -\dfrac{5}{3} & 
\cofg{4}{162} & = -\dfrac{3}{2} & 
\cofg{4}{163} & = \dfrac{17}{6} - 3 \zeta_3 & 
\cofg{4}{164} & = -6 \\ 
\cofg{4}{165} & = -1 & 
\cofg{4}{166} & = -1 & 
\cofg{4}{167} & = -\dfrac{1}{2} & 
\cofg{4}{168} & = -\dfrac{1}{2} \\ 
\cofg{4}{169} & = -\dfrac{3}{2} & 
\cofg{4}{170} & = \dfrac{34}{9} - \dfrac{4}{3} \zeta_3 & 
\cofg{4}{171} & = -\dfrac{2}{9} - \dfrac{4}{3} \zeta_3 & 
\cofg{4}{172} & = \dfrac{4}{3} \zeta_3 - \dfrac{31}{9} \\ 
\cofg{4}{173} & = \dfrac{4}{3} \zeta_3 - \dfrac{17}{18} & 
\cofg{4}{174} & = -\dfrac{15}{8} & 
\cofg{4}{175} & = \dfrac{11}{18} + \dfrac{2}{3} \zeta_3 & 
\cofg{4}{176} & = -\dfrac{3}{2} \\ 
\cofg{4}{177} & = -\dfrac{113}{72} - \dfrac{1}{6} \zeta_3 & 
\cofg{4}{178} & = -\dfrac{55}{72} - \dfrac{1}{3} \zeta_3 & 
\cofg{4}{179} & = \dfrac{115}{72} - \dfrac{1}{6} \zeta_3 & 
\cofg{4}{180} & = \dfrac{9}{8} \\ 
\cofg{4}{181} & = \dfrac{11}{18} - \dfrac{1}{3} \zeta_3 & 
\cofg{4}{182} & = \dfrac{43}{36} - \dfrac{1}{3} \zeta_3 & 
\cofg{4}{183} & = \dfrac{2}{3} \zeta_3 - \dfrac{115}{72} & 
\cofg{4}{184} & = -\dfrac{41}{16} \\ 
\cofg{4}{185} & = -\dfrac{3}{8} & 
\cofg{4}{186} & = -\dfrac{1}{4} & 
\cofg{4}{187} & = \dfrac{19}{96} & 
\cofg{4}{188} & = -\dfrac{9}{32} \\ 
\cofg{4}{189} & = \dfrac{1}{12} & 
\cofg{4}{190} & = -\dfrac{1}{2} & 
\cofg{4}{191} & = \dfrac{3}{16} & 
\cofg{4}{192} & = -\dfrac{17}{96} \\ 
\cofg{4}{193} & = -\dfrac{25}{32} & 
\cofg{4}{194} & = -\dfrac{3}{16} & 
\cofg{4}{195} & = -\dfrac{1}{4} & 
\cofg{4}{196} & = \dfrac{3}{16} \\ 
\cofg{4}{197} & = \dfrac{1}{16} & 
\cofg{4}{198} & = \dfrac{1}{16} & 
\cofg{4}{199} & = -4 & 
\cofg{4}{200} & = -2 \\ 
\cofg{4}{201} & = \dfrac{4}{3} - 4 \zeta_3 & 
\cofg{4}{202} & = \dfrac{8}{3} - 8 \zeta_3\,.
\end{align*}
}
Here $ \zeta_3= 1.20206... $ is Riemann's zeta function evaluated at 3.

\subsection{Yukawa \bef}
The $ \gamma_5 $-even TSs of the 3-loop Yukawa \bef are  
{\small
\begin{align*}
\beta^{(3)}_{a} =& \\
\cofy{3}{1} &\widetilde{T}^A \widetilde{T}^B y_a T^C T^B T^A T^C  & 
+\cofy{3}{2} &[S_2(S,C_S)]_{AB} y_b \big[T_\Phi^B T_\Phi^A \big]_{ab}  \\ 
+\cofy{3}{3} &y_a T^A T^B [S_2(S,C_S)]_{BA}  & 
+\cofy{3}{4} &[S_2(F,C_F)]_{AB} y_b \big[T_\Phi^A T_\Phi^B \big]_{ab}  \\ 
+\cofy{3}{5} &y_a T^A T^B [S_2(F,C_F)]_{AB}  & 
+\cofy{3}{6} &[C_2(S)]_{ab} y_c \big[C_2(S) C_2(S) \big]_{bc}  \\ 
+\cofy{3}{7} &\widetilde{C}_2(F) y_b \big[C_2(S) C_2(S) \big]_{ba}  & 
+\cofy{3}{8} &[C_2(S)]_{ba} \widetilde{C}_2(F) y_b C_2(F)  \\ 
+\cofy{3}{9} &[C_2(S)]_{ba} \widetilde{C}_2(F) \widetilde{C}_2(F) y_b  & 
+\cofy{3}{10} &\widetilde{C}_2(F) y_a C_2(F) C_2(F)  \\ 
+\cofy{3}{11} &\widetilde{C}_2(F) \widetilde{C}_2(F) \widetilde{C}_2(F) y_a  & 
+\cofy{3}{12} &y_b \big[C_2(S,G) C_2(S) \big]_{ba}  \\ 
+\cofy{3}{13} &[C_2(S)]_{ba} [C_2(S,S)]_{cb} y_c  & 
+\cofy{3}{14} &[C_2(S)]_{ba} [C_2(S,F)]_{cb} y_c  \\ 
+\cofy{3}{15} &[C_2(S)]_{ba} \widetilde{C}_2(F,G) y_b  & 
+\cofy{3}{16} &[C_2(S)]_{ba} \widetilde{C}_2(F,S) y_b  \\ 
+\cofy{3}{17} &[C_2(S,G)]_{ba} \widetilde{C}_2(F) y_b  & 
+\cofy{3}{18} &[C_2(S,S)]_{ba} \widetilde{C}_2(F) y_b  \\ 
+\cofy{3}{19} &[C_2(S)]_{ba} \widetilde{C}_2(F,F) y_b  & 
+\cofy{3}{20} &\widetilde{C}_2(F) y_a C_2(F,G)  \\ 
+\cofy{3}{21} &\widetilde{C}_2(F) y_a C_2(F,S)  & 
+\cofy{3}{22} &\widetilde{C}_2(F) \widetilde{C}_2(F,G) y_a  \\ 
+\cofy{3}{23} &\widetilde{C}_2(F) \widetilde{C}_2(F,S) y_a  & 
+\cofy{3}{24} &[C_2(S,F)]_{ba} \widetilde{C}_2(F) y_b  \\ 
+\cofy{3}{25} &\widetilde{C}_2(F) y_a C_2(F,F)  & 
+\cofy{3}{26} &\widetilde{C}_2(F) \widetilde{C}_2(F,F) y_a  \\ 
+\cofy{3}{27} &[S_2(S)]_{AB} [S_2(S)]_{CB} y_b \big[T_\Phi^A T_\Phi^C \big]_{ab}  & 
+\cofy{3}{28} &[C_2(G)]_{AB} [S_2(S)]_{AC} y_b \big[T_\Phi^B T_\Phi^C \big]_{ab}  \\ 
+\cofy{3}{29} &[C_2(G)]_{AB} [C_2(G)]_{AC} y_b \big[T_\Phi^B T_\Phi^C \big]_{ab}  & 
+\cofy{3}{30} &[S_2(F)]_{AB} [S_2(S)]_{BC} y_b \big[T_\Phi^A T_\Phi^C \big]_{ab}  \\ 
+\cofy{3}{31} &[C_2(G)]_{AB} [S_2(F)]_{AC} y_b \big[T_\Phi^B T_\Phi^C \big]_{ab}  & 
+\cofy{3}{32} &y_a T^A T^B [S_2(S)]_{AC} [S_2(S)]_{BC}  \\ 
+\cofy{3}{33} &[C_2(G)]_{AB} y_a T^C T^B [S_2(S)]_{AC}  & 
+\cofy{3}{34} &[C_2(G)]_{AB} [C_2(G)]_{AC} y_a T^B T^C  \\ 
+\cofy{3}{35} &[S_2(F)]_{AB} [S_2(F)]_{AC} y_b \big[T_\Phi^B T_\Phi^C \big]_{ab}  & 
+\cofy{3}{36} &y_a T^A T^B [S_2(F)]_{BC} [S_2(S)]_{AC}  \\ 
+\cofy{3}{37} &[C_2(G)]_{AB} y_a T^C T^B [S_2(F)]_{AC}  & 
+\cofy{3}{38} &y_a T^A T^B [S_2(F)]_{AC} [S_2(F)]_{BC}  \\ 
+\cofy{3}{39} &\widetilde{T}^A \widetilde{T}^B y_b \lambda_{cdba} \big[T_\Phi^B T_\Phi^A \big]_{cd}  & 
+\cofy{3}{40} &\lambda_{bcda} y_e \big[T_\Phi^A T_\Phi^B \big]_{bc} \big[T_\Phi^A T_\Phi^B \big]_{de}  \\ 
+\cofy{3}{41} &y_b \big[\Lambda_2 C_2(S) \big]_{cb}  & 
+\cofy{3}{42} &[C_2(S)]_{bc} \lambda_{bdea} \lambda_{cfde} y_f  \\ 
+\cofy{3}{43} &\lambda_{bcde} \lambda_{cfeg} \lambda_{dfga} y_b  & 
+\cofy{3}{44} &\widetilde{T}^A y_b T^B T^A \tilde{y}_a \widetilde{T}^B y_b  \\ 
+\cofy{3}{45} &\widetilde{T}^A \widetilde{T}^B y_b \Tr{y_b T^B T^A \tilde{y}_a }  & 
+\cofy{3}{46} &\widetilde{T}^A \widetilde{T}^B y_a \tilde{y}_b \widetilde{T}^B \widetilde{T}^A y_b  \\ 
+\cofy{3}{47} &[S_2(F,Y_F)]_{AB} y_b \big[T_\Phi^A T_\Phi^B \big]_{ab}  & 
+\cofy{3}{48} &y_a T^A T^B [S_2(F,Y_F)]_{AB}  \\ 
+\cofy{3}{49} &[S_2(S,Y_S)]_{AB} y_b \big[T_\Phi^A T_\Phi^B \big]_{ab}  & 
+\cofy{3}{50} &y_a T^A T^B [S_2(S,Y_S)]_{AB}  \\ 
+\cofy{3}{51} &[C_2(S)]_{bc} \widetilde{T}^A y_a \tilde{y}_b \widetilde{T}^A y_c  & 
+\cofy{3}{52} &[C_2(S)]_{ba} \widetilde{T}^A y_b \tilde{y}_c \widetilde{T}^A y_c  \\ 
+\cofy{3}{53} &\widetilde{T}^A y_a C_2(F) \tilde{y}_b \widetilde{T}^A y_b  & 
+\cofy{3}{54} &\widetilde{T}^A y_a \tilde{y}_b \widetilde{T}^A \widetilde{C}_2(F) y_b  \\ 
+\cofy{3}{55} &\widetilde{C}_2(F) y_b T^A \tilde{y}_b y_a T^A  & 
+\cofy{3}{56} &\widetilde{C}_2(F) y_a T^A \tilde{y}_b \widetilde{T}^A y_b  \\ 
+\cofy{3}{57} &\widetilde{C}_2(F) \widetilde{T}^A y_a \tilde{y}_b \widetilde{T}^A y_b  & 
+\cofy{3}{58} &[C_2(S)]_{ba} Y_2(F) \widetilde{T}^A y_b T^A  \\ 
+\cofy{3}{59} &[C_2(S)]_{bc} \widetilde{T}^A y_a T^A \tilde{y}_c y_b  & 
+\cofy{3}{60} &\widetilde{T}^A y_a T^A \tilde{y}_b \widetilde{C}_2(F) y_b  \\ 
+\cofy{3}{61} &\widetilde{C}_2(F) \widetilde{T}^A y_a T^A \widetilde{Y}_2(F)  & 
+\cofy{3}{62} & (-1) \widetilde{C}_2(F) \widetilde{T}^A y_b \big[T_\Phi^A Y_2(S) \big]_{ba}  \\ 
+\cofy{3}{63} &y_b \tilde{y}_a y_c \big[C_2(S) C_2(S) \big]_{cb}  & 
+\cofy{3}{64} &y_b \tilde{y}_c y_b \big[C_2(S) C_2(S) \big]_{ca}  \\ 
+\cofy{3}{65} &[C_2(S)]_{ba} [C_2(S)]_{cd} y_d \tilde{y}_b y_c  & 
+\cofy{3}{66} &[C_2(S)]_{bc} y_c C_2(F) \tilde{y}_a y_b  \\ 
+\cofy{3}{67} &[C_2(S)]_{ba} y_c C_2(F) \tilde{y}_b y_c  & 
+\cofy{3}{68} &[C_2(S)]_{bc} \widetilde{C}_2(F) y_c \tilde{y}_a y_b  \\ 
+\cofy{3}{69} &[C_2(S)]_{ba} \widetilde{C}_2(F) y_c \tilde{y}_b y_c  & 
+\cofy{3}{70} &y_b C_2(F) \tilde{y}_a \widetilde{C}_2(F) y_b  \\ 
+\cofy{3}{71} &y_b C_2(F) C_2(F) \tilde{y}_a y_b  & 
+\cofy{3}{72} &\widetilde{C}_2(F) y_b \tilde{y}_a y_b C_2(F)  \\ 
+\cofy{3}{73} &\widetilde{C}_2(F) y_b C_2(F) \tilde{y}_a y_b  & 
+\cofy{3}{74} &\widetilde{C}_2(F) \widetilde{C}_2(F) y_b \tilde{y}_a y_b  \\ 
+\cofy{3}{75} &\widetilde{C}_2(F) y_b \tilde{y}_a \widetilde{C}_2(F) y_b  & 
+\cofy{3}{76} &[C_2(S)]_{ba} \Tr{y_b C_2(F) \tilde{y}_c } y_c  \\ 
+\cofy{3}{77} &\Tr{\tilde{y}_b \widetilde{C}_2(F) y_a C_2(F) } y_b  & 
+\cofy{3}{78} &\Tr{C_2(F) C_2(F) \tilde{y}_b y_a } y_b  \\ 
+\cofy{3}{79} &y_a \tilde{y}_b y_c \big[C_2(S) C_2(S) \big]_{bc}  & 
+\cofy{3}{80} &[C_2(S)]_{bc} y_a \tilde{y}_b \widetilde{C}_2(F) y_c  \\ 
+\cofy{3}{81} &[C_2(S)]_{ba} \widetilde{C}_2(F) Y_2(F) y_b  & 
+\cofy{3}{82} &[C_2(S)]_{bc} \widetilde{C}_2(F) y_c \tilde{y}_b y_a  \\ 
+\cofy{3}{83} &y_a \tilde{y}_b \widetilde{C}_2(F) \widetilde{C}_2(F) y_b  & 
+\cofy{3}{84} &\widetilde{C}_2(F) \widetilde{C}_2(F) Y_2(F) y_a  \\ 
+\cofy{3}{85} &\widetilde{C}_2(F) y_b C_2(F) \tilde{y}_b y_a  & 
+\cofy{3}{86} &\widetilde{C}_2(F) y_a C_2(F) \widetilde{Y}_2(F)  \\ 
+\cofy{3}{87} &y_b \big[C_2(S) Y_2(S) C_2(S) \big]_{bc}  & 
+\cofy{3}{88} &\widetilde{C}_2(F) y_b \big[C_2(S) Y_2(S) \big]_{ac}  \\ 
+\cofy{3}{89} &[C_2(G)]_{AB} \widetilde{T}^B y_a \tilde{y}_b \widetilde{T}^A y_b  & 
+\cofy{3}{90} &\widetilde{T}^A y_a \tilde{y}_b \widetilde{T}^B y_b [S_2(S)]_{BA}  \\ 
+\cofy{3}{91} &\widetilde{T}^A y_a \tilde{y}_b \widetilde{T}^B y_b [S_2(F)]_{BA}  & 
+\cofy{3}{92} &[C_2(G)]_{AB} Y_2(F) \widetilde{T}^A y_a T^B  \\ 
+\cofy{3}{93} &Y_2(F) \widetilde{T}^A y_a T^B [S_2(S)]_{AB}  & 
+\cofy{3}{94} &Y_2(F) \widetilde{T}^A y_a T^B [S_2(F)]_{AB}  \\ 
+\cofy{3}{95} &[C_2(S,G)]_{bc} y_c \tilde{y}_a y_b  & 
+\cofy{3}{96} &[C_2(S,S)]_{bc} y_c \tilde{y}_a y_b  \\ 
+\cofy{3}{97} &[C_2(S,G)]_{ba} y_c \tilde{y}_b y_c  & 
+\cofy{3}{98} &[C_2(S,S)]_{ba} y_c \tilde{y}_b y_c  \\ 
+\cofy{3}{99} &y_b C_2(F,G) \tilde{y}_a y_b  & 
+\cofy{3}{100} &[C_2(S,F)]_{bc} y_c \tilde{y}_a y_b  \\ 
+\cofy{3}{101} &y_b C_2(F,S) \tilde{y}_a y_b  & 
+\cofy{3}{102} &[C_2(S,F)]_{ba} y_c \tilde{y}_b y_c  \\ 
+\cofy{3}{103} &\widetilde{C}_2(F,G) y_b \tilde{y}_a y_b  & 
+\cofy{3}{104} &\widetilde{C}_2(F,S) y_b \tilde{y}_a y_b  \\ 
+\cofy{3}{105} &y_b C_2(F,F) \tilde{y}_a y_b  & 
+\cofy{3}{106} &\widetilde{C}_2(F,F) y_b \tilde{y}_a y_b  \\ 
+\cofy{3}{107} &\Tr{y_b C_2(F,G) \tilde{y}_a } y_b  & 
+\cofy{3}{108} &\Tr{y_b C_2(F,S) \tilde{y}_a } y_b  \\ 
+\cofy{3}{109} &\Tr{y_b C_2(F,F) \tilde{y}_a } y_b  & 
+\cofy{3}{110} &[C_2(S,G)]_{bc} y_a \tilde{y}_b y_c  \\ 
+\cofy{3}{111} &[C_2(S,S)]_{bc} y_a \tilde{y}_b y_c  & 
+\cofy{3}{112} &y_a \tilde{y}_b \widetilde{C}_2(F,G) y_b  \\ 
+\cofy{3}{113} &y_a \tilde{y}_b \widetilde{C}_2(F,S) y_b  & 
+\cofy{3}{114} &[C_2(S,F)]_{bc} y_a \tilde{y}_b y_c  \\ 
+\cofy{3}{115} &y_a \widetilde{Y}_2(F) C_2(F,G)  & 
+\cofy{3}{116} &y_a \widetilde{Y}_2(F) C_2(F,S)  \\ 
+\cofy{3}{117} &y_a \tilde{y}_b \widetilde{C}_2(F,F) y_b  & 
+\cofy{3}{118} &y_a \widetilde{Y}_2(F) C_2(F,F)  \\ 
+\cofy{3}{119} &y_b \big[C_2(S,G) Y_2(S) \big]_{ba}  & 
+\cofy{3}{120} &[C_2(S,S)]_{bc} [Y_2(S)]_{ca} y_b  \\ 
+\cofy{3}{121} &[C_2(S,F)]_{bc} [Y_2(S)]_{ca} y_b  & 
+\cofy{3}{122} &\widetilde{T}^A y_b \tilde{y}_c \widetilde{T}^A y_d \lambda_{cdba}  \\ 
+\cofy{3}{123} &[C_2(S)]_{bc} y_d \tilde{y}_b y_e \lambda_{cdea}  & 
+\cofy{3}{124} &[C_2(S)]_{bc} y_b \tilde{y}_d y_e \lambda_{ceda}  \\ 
+\cofy{3}{125} &[C_2(S)]_{ab} y_c \tilde{y}_d y_e \lambda_{bdce}  & 
+\cofy{3}{126} &y_b C_2(F) \tilde{y}_c y_d \lambda_{cbda}  \\ 
+\cofy{3}{127} &\widetilde{C}_2(F) y_b \tilde{y}_c y_d \lambda_{bdca}  & 
+\cofy{3}{128} &y_b \tilde{y}_c y_d \lambda_{cefa} \lambda_{ebdf}  \\ 
+\cofy{3}{129} &y_b \tilde{y}_a y_c [\Lambda_2]_{cb}  & 
+\cofy{3}{130} &y_a \tilde{y}_b y_c [\Lambda_2]_{bc}  \\ 
+\cofy{3}{131} &y_b \tilde{y}_c y_d \lambda_{cdef} \lambda_{ebfa}  & 
+\cofy{3}{132} &\lambda_{bcda} \lambda_{cefd} [Y_2(S)]_{bf} y_e  \\ 
+\cofy{3}{133} &\widetilde{T}^A y_b \tilde{y}_c \widetilde{T}^A y_a \tilde{y}_b y_c  & 
+\cofy{3}{134} &[C_2(S)]_{bc} y_d \tilde{y}_c y_a \tilde{y}_d y_b  \\ 
+\cofy{3}{135} &[C_2(S)]_{ba} y_c \tilde{y}_d y_b \tilde{y}_c y_d  & 
+\cofy{3}{136} &y_b C_2(F) \tilde{y}_c y_a \tilde{y}_b y_c  \\ 
+\cofy{3}{137} &y_b \tilde{y}_c \widetilde{C}_2(F) y_a \tilde{y}_b y_c  & 
+\cofy{3}{138} &\widetilde{C}_2(F) y_b \tilde{y}_c y_a \tilde{y}_b y_c  \\ 
+\cofy{3}{139} &y_b T^A \tilde{y}_b y_a \tilde{y}_c \widetilde{T}^A y_c  & 
+\cofy{3}{140} &y_b T^A \tilde{y}_a y_c T^A \tilde{y}_c y_b  \\ 
+\cofy{3}{141} &y_b T^A \tilde{y}_b y_c T^A \tilde{y}_a y_c  & 
+\cofy{3}{142} &y_a \tilde{y}_b \widetilde{T}^A y_b \tilde{y}_c \widetilde{T}^A y_c  \\ 
+\cofy{3}{143} &\widetilde{T}^A y_b \tilde{y}_c \widetilde{T}^A y_c \tilde{y}_a y_b  & 
+\cofy{3}{144} &\widetilde{T}^A y_b \tilde{y}_a y_c \tilde{y}_b \widetilde{T}^A y_c  \\ 
+\cofy{3}{145} &\widetilde{T}^A y_b \tilde{y}_a y_b \tilde{y}_c \widetilde{T}^A y_c  & 
+\cofy{3}{146} &\widetilde{T}^A y_b T^A \tilde{y}_b y_c \tilde{y}_a y_c  \\ 
+\cofy{3}{147} &\widetilde{T}^A y_b \Tr{y_c T^A \tilde{y}_c y_b \tilde{y}_a }  & 
+\cofy{3}{148} &\widetilde{T}^A y_b \tilde{y}_c \widetilde{T}^A y_a \tilde{y}_c y_b  \\ 
+\cofy{3}{149} &\widetilde{T}^A y_a \tilde{y}_b y_c T^A \tilde{y}_c y_b  & 
+\cofy{3}{150} &\widetilde{T}^A y_a \tilde{y}_b \widetilde{T}^A y_c \tilde{y}_b y_c  \\ 
+\cofy{3}{151} &\widetilde{T}^A y_a T^A \tilde{y}_b y_c \tilde{y}_b y_c  & 
+\cofy{3}{152} &\widetilde{T}^A y_a \tilde{y}_b y_c \tilde{y}_b \widetilde{T}^A y_c  \\ 
+\cofy{3}{153} &y_a \tilde{y}_b y_c T^A \tilde{y}_c y_b T^A  & 
+\cofy{3}{154} & (-1) \widetilde{T}^A Y_2(F) y_a \tilde{y}_b \widetilde{T}^A y_b  \\ 
+\cofy{3}{155} &\widetilde{T}^A y_a \widetilde{Y}_2(F) \tilde{y}_b \widetilde{T}^A y_b  & 
+\cofy{3}{156} &\widetilde{T}^A y_a T^A \tilde{y}_b Y_2(F) y_b  \\ 
+\cofy{3}{157} &\widetilde{T}^A y_a \tilde{y}_b \widetilde{T}^A Y_2(F) y_b  & 
+\cofy{3}{158} & (-1) \widetilde{T}^A Y_2(F) y_b T^A \tilde{y}_a y_b  \\ 
+\cofy{3}{159} &y_a \widetilde{Y}_2(F) \tilde{y}_b \widetilde{T}^A y_b T^A  & 
+\cofy{3}{160} & (-1) y_a \tilde{y}_b \widetilde{T}^A y_b \widetilde{Y}_2(F) T^A  \\ 
+\cofy{3}{161} &Y_2(F) \widetilde{T}^A y_b \tilde{y}_a y_b T^A  & 
+\cofy{3}{162} &y_b T^A \tilde{y}_b y_c \big[T_\Phi^A Y_2(S) \big]_{ca}  \\ 
+\cofy{3}{163} &\widetilde{T}^A y_a T^A \tilde{y}_b y_c [Y_2(S)]_{cb}  & 
+\cofy{3}{164} &\widetilde{T}^A y_a \tilde{y}_b \widetilde{T}^A y_c [Y_2(S)]_{bc}  \\ 
+\cofy{3}{165} &Y_2(F) \widetilde{T}^A Y_2(F) y_a T^A  & 
+\cofy{3}{166} &Y_2(F) \widetilde{T}^A y_a T^A \widetilde{Y}_2(F)  \\ 
+\cofy{3}{167} & (-1) \widetilde{T}^A Y_2(F) y_b \big[T_\Phi^A Y_2(S) \big]_{ba}  & 
+\cofy{3}{168} &[C_2(S)]_{bc} y_d \tilde{y}_b y_a \tilde{y}_c y_d  \\ 
+\cofy{3}{169} &[C_2(S)]_{bc} y_d \tilde{y}_b y_d \tilde{y}_a y_c  & 
+\cofy{3}{170} &[C_2(S)]_{bc} y_d \tilde{y}_a y_b \tilde{y}_d y_c  \\ 
+\cofy{3}{171} &[C_2(S)]_{bc} y_c \tilde{y}_d y_a \tilde{y}_d y_b  & 
+\cofy{3}{172} &[C_2(S)]_{bc} y_a \tilde{y}_b y_d \tilde{y}_c y_d  \\ 
+\cofy{3}{173} &[C_2(S)]_{bc} y_a \tilde{y}_d y_b \tilde{y}_d y_c  & 
+\cofy{3}{174} &[C_2(S)]_{ba} y_c \tilde{y}_b y_d \tilde{y}_c y_d  \\ 
+\cofy{3}{175} &[C_2(S)]_{ba} y_c \tilde{y}_d y_b \tilde{y}_d y_c  & 
+\cofy{3}{176} &[C_2(S)]_{bc} \Tr{\tilde{y}_b y_d \tilde{y}_c y_a } y_d  \\ 
+\cofy{3}{177} &[C_2(S)]_{ba} \Tr{\tilde{y}_b y_c \tilde{y}_d y_c } y_d  & 
+\cofy{3}{178} &y_b \tilde{y}_c \widetilde{C}_2(F) y_b \tilde{y}_a y_c  \\ 
+\cofy{3}{179} &y_b C_2(F) \tilde{y}_c y_b \tilde{y}_a y_c  & 
+\cofy{3}{180} &y_b C_2(F) \tilde{y}_c y_a \tilde{y}_c y_b  \\ 
+\cofy{3}{181} &y_b \tilde{y}_a \widetilde{C}_2(F) y_c \tilde{y}_b y_c  & 
+\cofy{3}{182} &y_b \tilde{y}_c y_a C_2(F) \tilde{y}_c y_b  \\ 
+\cofy{3}{183} &y_b C_2(F) \tilde{y}_a y_c \tilde{y}_b y_c  & 
+\cofy{3}{184} &y_a \tilde{y}_b y_c \tilde{y}_b \widetilde{C}_2(F) y_c  \\ 
+\cofy{3}{185} &y_a \tilde{y}_b y_c C_2(F) \tilde{y}_b y_c  & 
+\cofy{3}{186} &y_a \tilde{y}_b \widetilde{C}_2(F) y_c \tilde{y}_b y_c  \\ 
+\cofy{3}{187} &\Tr{y_b C_2(F) \tilde{y}_c y_a \tilde{y}_c } y_b  & 
+\cofy{3}{188} &\Tr{y_a C_2(F) \tilde{y}_b y_c \tilde{y}_b } y_c  \\ 
+\cofy{3}{189} &\widetilde{C}_2(F) y_b \tilde{y}_a y_c \tilde{y}_b y_c  & 
+\cofy{3}{190} &\widetilde{C}_2(F) y_b \tilde{y}_c y_b \tilde{y}_a y_c  \\ 
+\cofy{3}{191} &\widetilde{C}_2(F) y_b \tilde{y}_c y_a \tilde{y}_c y_b  & 
+\cofy{3}{192} &\widetilde{C}_2(F) y_b \tilde{y}_c y_b \tilde{y}_c y_a  \\ 
+\cofy{3}{193} &y_b \tilde{y}_a y_c \Tr{y_b C_2(F) \tilde{y}_c }  & 
+\cofy{3}{194} &y_a \tilde{y}_b y_c \Tr{y_c C_2(F) \tilde{y}_b }  \\ 
+\cofy{3}{195} &[C_2(S)]_{bc} y_c \tilde{y}_a Y_2(F) y_b  & 
+\cofy{3}{196} &[C_2(S)]_{bc} y_a \tilde{y}_b Y_2(F) y_c  \\ 
+\cofy{3}{197} &[C_2(S)]_{ba} y_c \tilde{y}_b Y_2(F) y_c  & 
+\cofy{3}{198} &[C_2(S)]_{ba} \Tr{\tilde{y}_b Y_2(F) y_c } y_c  \\ 
+\cofy{3}{199} &[C_2(S)]_{bc} y_d \tilde{y}_b y_c \tilde{y}_a y_d  & 
+\cofy{3}{200} &[C_2(S)]_{bc} y_a \tilde{y}_d y_c \tilde{y}_b y_d  \\ 
+\cofy{3}{201} &[C_2(S)]_{bc} \Tr{\tilde{y}_c y_d \tilde{y}_a y_b } y_d  & 
+\cofy{3}{202} &y_b C_2(F) \widetilde{Y}_2(F) \tilde{y}_a y_b  \\ 
+\cofy{3}{203} &y_b C_2(F) \tilde{y}_a Y_2(F) y_b  & 
+\cofy{3}{204} &y_a \tilde{y}_b \widetilde{C}_2(F) Y_2(F) y_b  \\ 
+\cofy{3}{205} &\Tr{\widetilde{Y}_2(F) C_2(F) \tilde{y}_a y_b } y_b  & 
+\cofy{3}{206} &\Tr{y_b C_2(F) \tilde{y}_a Y_2(F) } y_b  \\ 
+\cofy{3}{207} &y_b \tilde{y}_a y_c C_2(F) \tilde{y}_c y_b  & 
+\cofy{3}{208} &y_a \tilde{y}_b y_c C_2(F) \tilde{y}_c y_b  \\ 
+\cofy{3}{209} &\Tr{y_b C_2(F) \tilde{y}_b y_c \tilde{y}_a } y_c  & 
+\cofy{3}{210} &\widetilde{C}_2(F) y_b \widetilde{Y}_2(F) \tilde{y}_a y_b  \\ 
+\cofy{3}{211} &\widetilde{C}_2(F) y_b \tilde{y}_a Y_2(F) y_b  & 
+\cofy{3}{212} &\widetilde{C}_2(F) Y_2(F) y_b \tilde{y}_a y_b  \\ 
+\cofy{3}{213} &\widetilde{C}_2(F) y_b \widetilde{Y}_2(F) \tilde{y}_b y_a  & 
+\cofy{3}{214} &y_b \tilde{y}_a y_c \big[C_2(S) Y_2(S) \big]_{cd}  \\ 
+\cofy{3}{215} &y_a \tilde{y}_b y_c \big[C_2(S) Y_2(S) \big]_{bd}  & 
+\cofy{3}{216} &[C_2(S)]_{ba} y_c \tilde{y}_b y_d [Y_2(S)]_{dc}  \\ 
+\cofy{3}{217} &y_b \tilde{y}_c y_b \big[C_2(S) Y_2(S) \big]_{ad}  & 
+\cofy{3}{218} &y_b C_2(F) \tilde{y}_a y_c [Y_2(S)]_{cb}  \\ 
+\cofy{3}{219} &y_a \tilde{y}_b \widetilde{C}_2(F) y_c [Y_2(S)]_{bc}  & 
+\cofy{3}{220} &\widetilde{C}_2(F) y_b \tilde{y}_a y_c [Y_2(S)]_{cb}  \\ 
+\cofy{3}{221} &\widetilde{C}_2(F) y_b \tilde{y}_c y_a [Y_2(S)]_{cb}  & 
+\cofy{3}{222} &\widetilde{C}_2(F) Y_2(F) Y_2(F) y_a  \\ 
+\cofy{3}{223} &[Y_2(S)]_{ba} y_c \big[Y_2(S) C_2(S) \big]_{bc}  & 
+\cofy{3}{224} &y_b \tilde{y}_c y_d \tilde{y}_c y_e \lambda_{beda}  \\ 
+\cofy{3}{225} &y_b \tilde{y}_c y_b \tilde{y}_d y_e \lambda_{edca}  & 
+\cofy{3}{226} &y_b \tilde{y}_c y_d \tilde{y}_b y_e \lambda_{ceda}  \\ 
+\cofy{3}{227} &y_b \tilde{y}_c y_d \tilde{y}_e y_b \lambda_{deca}  & 
+\cofy{3}{228} &y_b \tilde{y}_c y_a \tilde{y}_d y_e \lambda_{bedc}  \\ 
+\cofy{3}{229} &y_b \tilde{y}_c y_d \tilde{y}_a y_e \lambda_{cbed}  & 
+\cofy{3}{230} &y_a \tilde{y}_b y_c \tilde{y}_d y_e \lambda_{cedb}  \\ 
+\cofy{3}{231} &\lambda_{bcda} \Tr{\tilde{y}_d y_e \tilde{y}_c y_b } y_e  & 
+\cofy{3}{232} &\lambda_{bcde} \Tr{\tilde{y}_e y_c \tilde{y}_d y_a } y_b  \\ 
+\cofy{3}{233} &y_b \tilde{y}_c Y_2(F) y_d \lambda_{bdca}  & 
+\cofy{3}{234} &y_b \tilde{y}_c y_d \lambda_{ebda} [Y_2(S)]_{ce}  \\ 
+\cofy{3}{235} &y_b \tilde{y}_c y_d \lambda_{edca} [Y_2(S)]_{be}  & 
+\cofy{3}{236} &y_b \tilde{y}_c y_d \tilde{y}_a y_d \tilde{y}_b y_c  \\ 
+\cofy{3}{237} &y_b \tilde{y}_c y_d \tilde{y}_a y_b \tilde{y}_d y_c  & 
+\cofy{3}{238} &y_b \tilde{y}_c y_a \tilde{y}_d y_b \tilde{y}_d y_c  \\ 
+\cofy{3}{239} &y_b \tilde{y}_c y_d \Tr{\tilde{y}_c y_d \tilde{y}_b y_a }  & 
+\cofy{3}{240} &y_b \tilde{y}_c y_b \tilde{y}_d y_a \tilde{y}_c y_d  \\ 
+\cofy{3}{241} &y_b \tilde{y}_c y_d \tilde{y}_a y_b \tilde{y}_c y_d  & 
+\cofy{3}{242} &y_b \tilde{y}_c y_d \tilde{y}_b y_a \tilde{y}_c y_d  \\ 
+\cofy{3}{243} &y_b \tilde{y}_a y_c \tilde{y}_d y_b \tilde{y}_c y_d  & 
+\cofy{3}{244} &y_b \tilde{y}_c y_d \tilde{y}_a y_c \tilde{y}_d y_b  \\ 
+\cofy{3}{245} &y_a \tilde{y}_b y_c \tilde{y}_d y_b \tilde{y}_c y_d  & 
+\cofy{3}{246} &\Tr{\tilde{y}_b y_c \tilde{y}_d y_a \tilde{y}_c y_d } y_b  \\ 
+\cofy{3}{247} &y_b \tilde{y}_c y_a \widetilde{Y}_2(F) \tilde{y}_b y_c  & 
+\cofy{3}{248} &y_b \tilde{y}_c y_a \tilde{y}_b Y_2(F) y_c  \\ 
+\cofy{3}{249} &y_b \tilde{y}_c y_a \tilde{y}_b y_d [Y_2(S)]_{dc}  & 
+\cofy{3}{250} &y_b \tilde{y}_c y_d \tilde{y}_b y_a \tilde{y}_d y_c  \\ 
+\cofy{3}{251} &y_b \tilde{y}_c y_b \tilde{y}_d y_a \tilde{y}_d y_c  & 
+\cofy{3}{252} &y_b \tilde{y}_c y_d \Tr{\tilde{y}_c y_d \tilde{y}_a y_b }  \\ 
+\cofy{3}{253} &y_b \tilde{y}_c y_b \tilde{y}_a y_d \tilde{y}_c y_d  & 
+\cofy{3}{254} &y_b \tilde{y}_c y_d \tilde{y}_b y_d \tilde{y}_a y_c  \\ 
+\cofy{3}{255} &y_b \tilde{y}_a y_c \tilde{y}_d y_b \tilde{y}_d y_c  & 
+\cofy{3}{256} &y_b \tilde{y}_a y_c \tilde{y}_b y_d \tilde{y}_c y_d  \\ 
+\cofy{3}{257} &y_b \tilde{y}_a y_c \Tr{\tilde{y}_d y_c \tilde{y}_d y_b }  & 
+\cofy{3}{258} &y_b \tilde{y}_c y_d \tilde{y}_c y_a \tilde{y}_d y_b  \\ 
+\cofy{3}{259} &y_b \tilde{y}_c y_d \tilde{y}_a y_d \tilde{y}_c y_b  & 
+\cofy{3}{260} &y_b \tilde{y}_a y_c \tilde{y}_d y_c \tilde{y}_d y_b  \\ 
+\cofy{3}{261} &y_a \tilde{y}_b y_c \tilde{y}_d y_c \tilde{y}_d y_b  & 
+\cofy{3}{262} &y_a \tilde{y}_b y_c \Tr{\tilde{y}_d y_b \tilde{y}_d y_c }  \\ 
+\cofy{3}{263} &y_a \tilde{y}_b y_c \tilde{y}_b y_d \tilde{y}_c y_d  & 
+\cofy{3}{264} &y_a \tilde{y}_b y_c \tilde{y}_d y_b \tilde{y}_d y_c  \\ 
+\cofy{3}{265} &y_a \tilde{y}_b y_c \tilde{y}_d y_c \tilde{y}_b y_d  & 
+\cofy{3}{266} &\Tr{\tilde{y}_b y_c \tilde{y}_d y_c \tilde{y}_d y_a } y_b  \\ 
+\cofy{3}{267} &\Tr{\tilde{y}_b y_c \tilde{y}_d y_c \tilde{y}_a y_d } y_b  & 
+\cofy{3}{268} &\Tr{\tilde{y}_b y_c \tilde{y}_d y_a \tilde{y}_d y_c } y_b  \\ 
+\cofy{3}{269} &y_b \tilde{y}_c Y_2(F) y_b \tilde{y}_a y_c  & 
+\cofy{3}{270} &y_b \tilde{y}_a y_c \Tr{\tilde{y}_b Y_2(F) y_c }  \\ 
+\cofy{3}{271} &y_b \tilde{y}_c Y_2(F) y_c \tilde{y}_a y_b  & 
+\cofy{3}{272} &y_b \tilde{y}_c y_b \widetilde{Y}_2(F) \tilde{y}_a y_c  \\ 
+\cofy{3}{273} &y_b \tilde{y}_c Y_2(F) y_a \tilde{y}_c y_b  & 
+\cofy{3}{274} &y_b \tilde{y}_a y_c \tilde{y}_b Y_2(F) y_c  \\ 
+\cofy{3}{275} &y_b \tilde{y}_c y_a \tilde{y}_c Y_2(F) y_b  & 
+\cofy{3}{276} &y_b \tilde{y}_c y_b \tilde{y}_a Y_2(F) y_c  \\ 
+\cofy{3}{277} &y_a \tilde{y}_b Y_2(F) y_c \tilde{y}_b y_c  & 
+\cofy{3}{278} &y_a \tilde{y}_b y_c \widetilde{Y}_2(F) \tilde{y}_b y_c  \\ 
+\cofy{3}{279} &y_a \tilde{y}_b y_c \Tr{\tilde{y}_c Y_2(F) y_b }  & 
+\cofy{3}{280} &y_a \tilde{y}_b y_c \widetilde{Y}_2(F) \tilde{y}_c y_b  \\ 
+\cofy{3}{281} &y_a \tilde{y}_b y_c \tilde{y}_b Y_2(F) y_c  & 
+\cofy{3}{282} &\Tr{\tilde{y}_b Y_2(F) y_c \tilde{y}_a y_c } y_b  \\ 
+\cofy{3}{283} &\Tr{\tilde{y}_b Y_2(F) y_b \tilde{y}_c y_a } y_c  & 
+\cofy{3}{284} &\Tr{\tilde{y}_b Y_2(F) y_a \tilde{y}_b y_c } y_c  \\ 
+\cofy{3}{285} &y_b \tilde{y}_c y_a \tilde{y}_d y_b [Y_2(S)]_{cd}  & 
+\cofy{3}{286} &y_b \tilde{y}_c y_d \tilde{y}_a y_b [Y_2(S)]_{cd}  \\ 
+\cofy{3}{287} &y_b \tilde{y}_c y_b \tilde{y}_a y_d [Y_2(S)]_{cd}  & 
+\cofy{3}{288} &y_b \tilde{y}_a y_c \tilde{y}_b y_d [Y_2(S)]_{cd}  \\ 
+\cofy{3}{289} &y_b \tilde{y}_c y_a \tilde{y}_c y_d [Y_2(S)]_{db}  & 
+\cofy{3}{290} &y_a \tilde{y}_b y_c \tilde{y}_d y_c [Y_2(S)]_{bd}  \\ 
+\cofy{3}{291} &y_a \tilde{y}_b y_c \tilde{y}_d y_b [Y_2(S)]_{dc}  & 
+\cofy{3}{292} &y_a \tilde{y}_b y_c \tilde{y}_b y_d [Y_2(S)]_{dc}  \\ 
+\cofy{3}{293} &\Tr{\tilde{y}_b y_c \tilde{y}_a y_d } [Y_2(S)]_{dc} y_b  & 
+\cofy{3}{294} &\Tr{\tilde{y}_b y_c \tilde{y}_d y_a } [Y_2(S)]_{bc} y_d  \\ 
+\cofy{3}{295} &y_b \widetilde{Y}_2(F) \tilde{y}_a Y_2(F) y_b  & 
+\cofy{3}{296} &y_b \tilde{y}_a Y_2(F) Y_2(F) y_b  \\ 
+\cofy{3}{297} &y_a \tilde{y}_b Y_2(F) Y_2(F) y_b  & 
+\cofy{3}{298} &\Tr{\tilde{y}_b Y_2(F) y_a \widetilde{Y}_2(F) } y_b  \\ 
+\cofy{3}{299} &\Tr{\tilde{y}_b Y_2(F) Y_2(F) y_a } y_b  & 
+\cofy{3}{300} &y_b \tilde{y}_a Y_2(F) y_c [Y_2(S)]_{cb}  \\ 
+\cofy{3}{301} &y_a \tilde{y}_b Y_2(F) y_c [Y_2(S)]_{bc}  & 
+\cofy{3}{302} &y_b \tilde{y}_a y_c \big[Y_2(S) Y_2(S) \big]_{bd}  \\ 
+\cofy{3}{303} &y_a \tilde{y}_b y_c \big[Y_2(S) Y_2(S) \big]_{bd} \, . 
\end{align*}
{\normalsize There are an additional 5 $ \gamma_5 $-odd TSs:}
\begin{align*}
+\cofy{3}{304} & \sigma_3 y_b T^A T^B \Tr{\sigma^3 y_b T^A T^B \tilde{y}_a} &
+\cofy{3}{305} & \sigma_3 \widetilde{T}^A y_b T^B \Tr{\sigma^3 \widetilde{T}^A y_b T^B \tilde{y}_a}\\
+\cofy{3}{306} & \sigma_3 y_b T^A T^B \Tr{\sigma^3 y_b T^B T^A \tilde{y}_a} &
+\cofy{3}{307} & \sigma_3 y_b T^A T^B \Tr{\sigma^3 \widetilde{T}^A y_b T^B \tilde{y}_a} \\
+\cofy{3}{308} & \sigma_3 \widetilde{T}^A y_b T^B \Tr{\sigma^3 y_b T^A T^B \tilde{y}_a} \,.
\end{align*}
}

The coefficients of the 3-loop Yukawa \bef are 
{\small
\begin{align*}
\cofy{3}{1} & = -72 & 
\cofy{3}{2} & = -3 & 
\cofy{3}{3} & = 33 - 12 \zeta_3 & 
\cofy{3}{4} & = -\dfrac{3}{2} \\ 
\cofy{3}{5} & = \dfrac{45}{2} - 24 \zeta_3 & 
\cofy{3}{6} & = \dfrac{93}{2} & 
\cofy{3}{7} & = -78 & 
\cofy{3}{8} & = -12 \\ 
\cofy{3}{9} & = 102 & 
\cofy{3}{10} & = -48 & 
\cofy{3}{11} & = -9 & 
\cofy{3}{12} & = -\dfrac{297}{2} \\ 
\cofy{3}{13} & = \dfrac{53}{8} & 
\cofy{3}{14} & = \dfrac{25}{4} & 
\cofy{3}{15} & = 62 & 
\cofy{3}{16} & = -3 \\ 
\cofy{3}{17} & = 48 & 
\cofy{3}{18} & = -4 & 
\cofy{3}{19} & = -2 & 
\cofy{3}{20} & = 102 \\ 
\cofy{3}{21} & = -3 & 
\cofy{3}{22} & = -\dfrac{219}{2} & 
\cofy{3}{23} & = \dfrac{17}{4} & 
\cofy{3}{24} & = -4 \\ 
\cofy{3}{25} & = -6 & 
\cofy{3}{26} & = \dfrac{13}{2} & 
\cofy{3}{27} & = \dfrac{11}{144} & 
\cofy{3}{28} & = -\dfrac{1165}{144} \\ 
\cofy{3}{29} & = \dfrac{9721}{72} & 
\cofy{3}{30} & = \dfrac{2}{9} & 
\cofy{3}{31} & = -\dfrac{925}{72} & 
\cofy{3}{32} & = \dfrac{5}{216} \\ 
\cofy{3}{33} & = \dfrac{181}{27} + 12 \zeta_3 & 
\cofy{3}{34} & = -\dfrac{10441}{54} & 
\cofy{3}{35} & = \dfrac{5}{36} & 
\cofy{3}{36} & = \dfrac{19}{27} \\ 
\cofy{3}{37} & = \dfrac{278}{27} + 24 \zeta_3 & 
\cofy{3}{38} & = \dfrac{35}{54} & 
\cofy{3}{39} & = -12 & 
\cofy{3}{40} & = \dfrac{5}{4} \\ 
\cofy{3}{41} & = -\dfrac{17}{48} & 
\cofy{3}{42} & = \dfrac{19}{16} & 
\cofy{3}{43} & = -\dfrac{1}{16} & 
\cofy{3}{44} & = -2 - 144 \zeta_3 \\ 
\cofy{3}{45} & = 36 - 72 \zeta_3 & 
\cofy{3}{46} & = 8 & 
\cofy{3}{47} & = \dfrac{3}{4} & 
\cofy{3}{48} & = -\dfrac{23}{4} \\ 
\cofy{3}{49} & = 0 & 
\cofy{3}{50} & = -\dfrac{9}{4} & 
\cofy{3}{51} & = 44 & 
\cofy{3}{52} & = -64 \\ 
\cofy{3}{53} & = 6 & 
\cofy{3}{54} & = 22 & 
\cofy{3}{55} & = 0 & 
\cofy{3}{56} & = 24 \\ 
\cofy{3}{57} & = 16 & 
\cofy{3}{58} & = -4 & 
\cofy{3}{59} & = 64 & 
\cofy{3}{60} & = -58 \\ 
\cofy{3}{61} & = 4 & 
\cofy{3}{62} & = 10 - 12 \zeta_3 & 
\cofy{3}{63} & = \dfrac{65}{4} - 72 \zeta_3 & 
\cofy{3}{64} & = 80 - 42 \zeta_3 \\ 
\cofy{3}{65} & = -\dfrac{59}{2} - 36 \zeta_3 & 
\cofy{3}{66} & = 72 \zeta_3 - 15 & 
\cofy{3}{67} & = 36 \zeta_3 - \dfrac{149}{2} & 
\cofy{3}{68} & = 16 + 144 \zeta_3 \\ 
\cofy{3}{69} & = 36 \zeta_3 - \dfrac{141}{2} & 
\cofy{3}{70} & = 72 \zeta_3 - \dfrac{47}{2} & 
\cofy{3}{71} & = -17 & 
\cofy{3}{72} & = 15 - 24 \zeta_3 \\ 
\cofy{3}{73} & = \dfrac{47}{2} - 36 \zeta_3 & 
\cofy{3}{74} & = 15 - 48 \zeta_3 & 
\cofy{3}{75} & = \dfrac{63}{2} - 36 \zeta_3 & 
\cofy{3}{76} & = -\dfrac{11}{2} \\ 
\cofy{3}{77} & = 9 \zeta_3 - \dfrac{11}{8} & 
\cofy{3}{78} & = 9 \zeta_3 - \dfrac{27}{2} & 
\cofy{3}{79} & = \dfrac{235}{8} - 15 \zeta_3 & 
\cofy{3}{80} & = 36 \zeta_3 - \dfrac{53}{4} \\ 
\cofy{3}{81} & = 5 & 
\cofy{3}{82} & = 36 \zeta_3 - \dfrac{297}{4} & 
\cofy{3}{83} & = 3 \zeta_3 - \dfrac{83}{4} & 
\cofy{3}{84} & = 17 - 33 \zeta_3 \\ 
\cofy{3}{85} & = 26 - 6 \zeta_3 & 
\cofy{3}{86} & = -5 & 
\cofy{3}{87} & = \dfrac{139}{16} - \dfrac{3}{2} \zeta_3 & 
\cofy{3}{88} & = 5 - 6 \zeta_3 \\ 
\cofy{3}{89} & = 81 & 
\cofy{3}{90} & = -3 & 
\cofy{3}{91} & = -2 & 
\cofy{3}{92} & = -19 \\ 
\cofy{3}{93} & = 0 & 
\cofy{3}{94} & = 0 & 
\cofy{3}{95} & = \dfrac{51}{8} & 
\cofy{3}{96} & = -\dfrac{11}{8} \\ 
\cofy{3}{97} & = 6 \zeta_3 - \dfrac{115}{2} & 
\cofy{3}{98} & = 3 & 
\cofy{3}{99} & = 99 - 36 \zeta_3 & 
\cofy{3}{100} & = -\dfrac{5}{4} \\ 
\cofy{3}{101} & = -\dfrac{25}{4} & 
\cofy{3}{102} & = 2 & 
\cofy{3}{103} & = \dfrac{61}{2} & 
\cofy{3}{104} & = -\dfrac{3}{2} \\ 
\cofy{3}{105} & = -\dfrac{11}{2} & 
\cofy{3}{106} & = -1 & 
\cofy{3}{107} & = \dfrac{77}{4} - 9 \zeta_3 & 
\cofy{3}{108} & = -\dfrac{11}{8} \\ 
\cofy{3}{109} & = -1 & 
\cofy{3}{110} & = \dfrac{785}{16} - 3 \zeta_3 & 
\cofy{3}{111} & = -\dfrac{43}{16} & 
\cofy{3}{112} & = -9 \zeta_3 \\ 
\cofy{3}{113} & = -\dfrac{3}{4} & 
\cofy{3}{114} & = -\dfrac{21}{8} & 
\cofy{3}{115} & = -\dfrac{9}{4} - 9 \zeta_3 & 
\cofy{3}{116} & = \dfrac{1}{8} \\ 
\cofy{3}{117} & = -\dfrac{3}{4} & 
\cofy{3}{118} & = 0 & 
\cofy{3}{119} & = \dfrac{3}{2} \zeta_3 - \dfrac{75}{32} & 
\cofy{3}{120} & = \dfrac{3}{32} \\ 
\cofy{3}{121} & = \dfrac{1}{16} & 
\cofy{3}{122} & = 0 & 
\cofy{3}{123} & = -\dfrac{17}{2} & 
\cofy{3}{124} & = -17 \\ 
\cofy{3}{125} & = \dfrac{19}{2} & 
\cofy{3}{126} & = 2 & 
\cofy{3}{127} & = 0 & 
\cofy{3}{128} & = \dfrac{3}{2} \\ 
\cofy{3}{129} & = -\dfrac{3}{8} & 
\cofy{3}{130} & = -\dfrac{11}{48} & 
\cofy{3}{131} & = 1 & 
\cofy{3}{132} & = -\dfrac{5}{32} \\ 
\cofy{3}{133} & = -30 & 
\cofy{3}{134} & = 6 & 
\cofy{3}{135} & = 0 & 
\cofy{3}{136} & = 10 \\ 
\cofy{3}{137} & = 7 & 
\cofy{3}{138} & = -3 & 
\cofy{3}{139} & = -6 & 
\cofy{3}{140} & = -26 \\ 
\cofy{3}{141} & = -14 & 
\cofy{3}{142} & = 2 & 
\cofy{3}{143} & = -14 & 
\cofy{3}{144} & = -12 \\ 
\cofy{3}{145} & = -22 & 
\cofy{3}{146} & = 12 & 
\cofy{3}{147} & = -8 & 
\cofy{3}{148} & = -6 \\ 
\cofy{3}{149} & = -12 & 
\cofy{3}{150} & = -22 & 
\cofy{3}{151} & = 24 & 
\cofy{3}{152} & = -22 \\ 
\cofy{3}{153} & = 2 & 
\cofy{3}{154} & = 0 & 
\cofy{3}{155} & = -5 & 
\cofy{3}{156} & = 3 \\ 
\cofy{3}{157} & = -5 & 
\cofy{3}{158} & = 0 & 
\cofy{3}{159} & = -\dfrac{3}{2} & 
\cofy{3}{160} & = \dfrac{3}{2} \\ 
\cofy{3}{161} & = 0 & 
\cofy{3}{162} & = 0 & 
\cofy{3}{163} & = 3 & 
\cofy{3}{164} & = -11 \\ 
\cofy{3}{165} & = 0 & 
\cofy{3}{166} & = 0 & 
\cofy{3}{167} & = 0 & 
\cofy{3}{168} & = -10 \\ 
\cofy{3}{169} & = 34 - 48 \zeta_3 & 
\cofy{3}{170} & = -8 & 
\cofy{3}{171} & = 4 & 
\cofy{3}{172} & = \dfrac{7}{4} - 6 \zeta_3 \\ 
\cofy{3}{173} & = \dfrac{7}{4} - 6 \zeta_3 & 
\cofy{3}{174} & = 2 & 
\cofy{3}{175} & = 36 & 
\cofy{3}{176} & = -5 \\ 
\cofy{3}{177} & = \dfrac{11}{2} & 
\cofy{3}{178} & = 24 \zeta_3 - 11 & 
\cofy{3}{179} & = 24 \zeta_3 - 17 & 
\cofy{3}{180} & = -24 \zeta_3 \\ 
\cofy{3}{181} & = 12 - 12 \zeta_3 & 
\cofy{3}{182} & = 24 \zeta_3 - 24 & 
\cofy{3}{183} & = 13 - 12 \zeta_3 & 
\cofy{3}{184} & = \dfrac{35}{8} - 3 \zeta_3 \\ 
\cofy{3}{185} & = 9 \zeta_3 - 3 & 
\cofy{3}{186} & = \dfrac{35}{8} - 3 \zeta_3 & 
\cofy{3}{187} & = -\dfrac{13}{8} & 
\cofy{3}{188} & = -\dfrac{13}{8} \\ 
\cofy{3}{189} & = 12 \zeta_3 - 3 & 
\cofy{3}{190} & = 12 \zeta_3 - 10 & 
\cofy{3}{191} & = -14 & 
\cofy{3}{192} & = 9 \zeta_3 - \dfrac{31}{4} \\ 
\cofy{3}{193} & = 12 \zeta_3 - \dfrac{51}{4} & 
\cofy{3}{194} & = 6 \zeta_3 - 8 & 
\cofy{3}{195} & = 1 & 
\cofy{3}{196} & = -\dfrac{1}{8} - 3 \zeta_3 \\ 
\cofy{3}{197} & = 12 + 12 \zeta_3 & 
\cofy{3}{198} & = \dfrac{5}{4} + 3 \zeta_3 & 
\cofy{3}{199} & = 12 \zeta_3 - \dfrac{77}{2} & 
\cofy{3}{200} & = 3 \zeta_3 - \dfrac{47}{8} \\ 
\cofy{3}{201} & = 3 \zeta_3 - \dfrac{67}{8} & 
\cofy{3}{202} & = \dfrac{87}{4} - 24 \zeta_3 & 
\cofy{3}{203} & = \dfrac{13}{2} - 12 \zeta_3 & 
\cofy{3}{204} & = \dfrac{25}{4} - 6 \zeta_3 \\ 
\cofy{3}{205} & = \dfrac{25}{8} - 6 \zeta_3 & 
\cofy{3}{206} & = \dfrac{5}{8} - 3 \zeta_3 & 
\cofy{3}{207} & = \dfrac{7}{4} + 12 \zeta_3 & 
\cofy{3}{208} & = 3 \zeta_3 - 1 \\ 
\cofy{3}{209} & = 3 \zeta_3 - \dfrac{25}{8} & 
\cofy{3}{210} & = -\dfrac{11}{2} & 
\cofy{3}{211} & = -\dfrac{11}{2} & 
\cofy{3}{212} & = 0 \\ 
\cofy{3}{213} & = 3 \zeta_3 - \dfrac{1}{4} & 
\cofy{3}{214} & = \dfrac{3}{2} - 12 \zeta_3 & 
\cofy{3}{215} & = -\dfrac{29}{8} - 6 \zeta_3 & 
\cofy{3}{216} & = 5 \\ 
\cofy{3}{217} & = 0 & 
\cofy{3}{218} & = \dfrac{15}{2} - 6 \zeta_3 & 
\cofy{3}{219} & = \dfrac{21}{4} - 3 \zeta_3 & 
\cofy{3}{220} & = 6 \zeta_3 - \dfrac{9}{2} \\ 
\cofy{3}{221} & = \dfrac{27}{8} + 3 \zeta_3 & 
\cofy{3}{222} & = 0 & 
\cofy{3}{223} & = -\dfrac{3}{4} & 
\cofy{3}{224} & = 3 \\ 
\cofy{3}{225} & = -2 & 
\cofy{3}{226} & = 4 & 
\cofy{3}{227} & = 5 & 
\cofy{3}{228} & = 2 \\ 
\cofy{3}{229} & = 6 & 
\cofy{3}{230} & = 2 & 
\cofy{3}{231} & = \dfrac{5}{8} & 
\cofy{3}{232} & = \dfrac{5}{8} \\ 
\cofy{3}{233} & = 1 & 
\cofy{3}{234} & = \dfrac{3}{2} & 
\cofy{3}{235} & = 1 & 
\cofy{3}{236} & = -3 \\ 
\cofy{3}{237} & = 12 \zeta_3 - 4 & 
\cofy{3}{238} & = 12 \zeta_3 - 6 & 
\cofy{3}{239} & = 12 \zeta_3 - 8 & 
\cofy{3}{240} & = 2 \\ 
\cofy{3}{241} & = 2 & 
\cofy{3}{242} & = 12 \zeta_3 - 4 & 
\cofy{3}{243} & = 12 \zeta_3 - 6 & 
\cofy{3}{244} & = 6 \zeta_3 - 5 \\ 
\cofy{3}{245} & = 3 \zeta_3 - 2 & 
\cofy{3}{246} & = \dfrac{3}{2} \zeta_3 - 1 & 
\cofy{3}{247} & = -3 & 
\cofy{3}{248} & = -1 \\ 
\cofy{3}{249} & = -1 & 
\cofy{3}{250} & = -8 & 
\cofy{3}{251} & = -6 & 
\cofy{3}{252} & = 0 \\ 
\cofy{3}{253} & = -2 & 
\cofy{3}{254} & = -4 & 
\cofy{3}{255} & = 2 & 
\cofy{3}{256} & = 4 \\ 
\cofy{3}{257} & = \dfrac{5}{4} & 
\cofy{3}{258} & = -2 & 
\cofy{3}{259} & = 4 & 
\cofy{3}{260} & = -2 \\ 
\cofy{3}{261} & = -\dfrac{5}{8} & 
\cofy{3}{262} & = 1 & 
\cofy{3}{263} & = 1 & 
\cofy{3}{264} & = 0 \\ 
\cofy{3}{265} & = 0 & 
\cofy{3}{266} & = -\dfrac{3}{8} & 
\cofy{3}{267} & = -\dfrac{3}{4} & 
\cofy{3}{268} & = \dfrac{7}{4} \\ 
\cofy{3}{269} & = 1 & 
\cofy{3}{270} & = \dfrac{25}{8} & 
\cofy{3}{271} & = \dfrac{7}{8} & 
\cofy{3}{272} & = -3 \\ 
\cofy{3}{273} & = 4 & 
\cofy{3}{274} & = 3 & 
\cofy{3}{275} & = -2 & 
\cofy{3}{276} & = -3 \\ 
\cofy{3}{277} & = \dfrac{3}{16} & 
\cofy{3}{278} & = \dfrac{1}{2} & 
\cofy{3}{279} & = 2 & 
\cofy{3}{280} & = \dfrac{1}{8} \\ 
\cofy{3}{281} & = \dfrac{3}{16} & 
\cofy{3}{282} & = \dfrac{7}{16} & 
\cofy{3}{283} & = \dfrac{5}{16} & 
\cofy{3}{284} & = \dfrac{7}{16} \\ 
\cofy{3}{285} & = 2 & 
\cofy{3}{286} & = \dfrac{25}{8} & 
\cofy{3}{287} & = -3 & 
\cofy{3}{288} & = 3 \\ 
\cofy{3}{289} & = -1 & 
\cofy{3}{290} & = -\dfrac{3}{16} & 
\cofy{3}{291} & = \dfrac{9}{16} & 
\cofy{3}{292} & = -\dfrac{3}{16} \\ 
\cofy{3}{293} & = \dfrac{9}{16} & 
\cofy{3}{294} & = 1 & 
\cofy{3}{295} & = -\dfrac{1}{2} & 
\cofy{3}{296} & = -1 \\ 
\cofy{3}{297} & = -\dfrac{5}{16} & 
\cofy{3}{298} & = \dfrac{1}{32} & 
\cofy{3}{299} & = -\dfrac{3}{16} & 
\cofy{3}{300} & = -1 \\ 
\cofy{3}{301} & = -\dfrac{1}{4} & 
\cofy{3}{302} & = -\dfrac{1}{2} & 
\cofy{3}{303} & = -\dfrac{3}{16} & 
\cofy{3}{304} & = -24 \\ 
\cofy{3}{305} & = -12 & 
\cofy{3}{306} & = 8 - 24 \zeta_3 & 
\cofy{3}{307} & = 8 - 24 \zeta_3 & 
\cofy{3}{308} & = 8 - 24 \zeta_3\,.
\end{align*}
}

\subsection{$ \upsilon $-function} 
The fermion representation, $ \upsilon^{(3)}\du{i}{j} $, of the 3-loop $ \upsilon $-function is parametrized with TSs
\begin{align*}
F^{(3)}_1 =& \widetilde{T}^A Y_2(F) y_b T^A \tilde{y}_b & 
F^{(3)}_2 =& [C_2(S)]_{ab} y_a \tilde{y}_c y_b \tilde{y}_c \\ 
F^{(3)}_3 =&  y_a C_2(F) \tilde{y}_b y_a \tilde{y}_b & 
F^{(3)}_4 =& y_a \tilde{y}_b y_c \tilde{y}_b y_a \tilde{y}_c \\
F^{(3)}_5 =& y_a \widetilde{Y}_2(F) \tilde{y}_b y_a \tilde{y}_b &
F^{(3)}_6 =& [Y_2(S)]_{ab} y_a \tilde{y}_c y_b \tilde{y}_c\,.
\end{align*}
The TSs for the scalar representation, $ \upsilon^{(3)}_{ab} $, are given by 
\begin{align*}
[S^{(3)}_1]_{ab} =& \Tr{y_a C_2(F) \tilde{y}_c y_b \tilde{y}_c} & 
[S^{(3)}_2]_{ab} =& \Tr{y_a \tilde{y}_c y_d \tilde{y}_e} \lambda_{cdeb} \\ 
[S^{(3)}_3]_{ab} =& \Tr{y_a \widetilde{Y}_2(F) \tilde{y}_c y_b \tilde{y}_c} \,.
\end{align*}

The coefficients of the 3-loop $ \upsilon $-function are 
\begin{align*}
\coff{3}{1} & = 0 & 
\coff{3}{2} & = \dfrac{29}{8} - 3 \zeta_3 & 
\coff{3}{3} & = \dfrac{21}{8} - 3 \zeta_3 \\ 
\coff{3}{4} & = -\dfrac{3}{8} & 
\coff{3}{5} & = -\dfrac{5}{16} & 
\coff{3}{6} & = -\dfrac{7}{16} \\ 
\cofs{3}{1} & = \dfrac{7}{2} - 6 \zeta_3 & 
\cofs{3}{2} & = \dfrac{5}{8} & 
\cofs{3}{3} & = -\dfrac{3}{4}\,.
\end{align*}

\sectionlike{References}
\vspace{-10pt}
\bibliography{References}

\end{document}

%% file: PreRamble.tex
\usepackage[english]{babel}
\usepackage{microtype,xspace}
\usepackage[utf8]{inputenc}	% Allows for writing special charachters in the tex-file 

\usepackage{amsfonts,amsmath,amssymb,bm,mathrsfs,mathtools,dsfont} 	% Standard mathematics 
\allowdisplaybreaks 	%Allows pagebreak during multiline math enviroments
\usepackage[table]{xcolor}
\usepackage{hyperref}
\usepackage{slashed}
\usepackage{multicol}

\usepackage{siunitx}
\usepackage{geometry}
\usepackage[sort&compress,numbers,merge]{natbib}
\usepackage{chapterbib}
\addto\captionsenglish{\renewcommand*{\bibname}{References}}
\makeatletter
\renewcommand{\@memb@bchap}{ %\section*{\bibname} 
\bibmark \prebibhook
}
\makeatother

\usepackage{graphicx}
\graphicspath{{./Figures/}}
\usepackage{caption,subcaption}
\usepackage{multirow}

\usepackage{tabularx}
\newcolumntype{Y}{>{\centering\arraybackslash}X}

\usepackage[shortlabels]{enumitem}
\setlist{itemsep=.1em,topsep=.5em}
\SetEnumerateShortLabel{i}{\textit{\roman*}}

\definecolor{red}{rgb}{0.6,.0706,.1373}
\definecolor{blue}{rgb}{0,0.396,0.741}
\colorlet{blueRef}{blue!80!black}
\colorlet{blueLink}{blue!100!black}
\hypersetup{
	colorlinks, 
	bookmarksopen, 
	bookmarksnumbered,
	citecolor=blueLink, 	%color of links to bibliography
	linkcolor=blueLink,	%color of internal links
	urlcolor=blueLink,			%color of external links 
	pdftitle={4-3-2 Beta functions},    %   - title (PDF meta)
	pdfsubject={Beta functions in particle physics},	%   - subject (PDF meta)
	pdfauthor={Joshua Davies, Florian Herren, and Anders Eller Thomsen},    	%   - author (PDF meta)
}

\addto\captionsenglish{}

\renewcommand{\contentsname}{Contents}
\renewcommand{\printtoctitle}[1]{}

\settocdepth{subsection}
\newcommand{\toc}{ {
	\hypersetup{linkcolor = black} %Locally changes the linkcolor for the toc
	\vspace*{-.06\textheight}	
	\tableofcontents*
	\thispagestyle{empty} 
} }

\newcommand{\app}[1][Appendices]{
	\renewcommand{\thesubsection}{\Alph{subsection}}
	\numberwithin{equation}{subsection}
	\numberwithin{figure}{subsection}
	\pagestyle{appstyle}
	\sectionlike{#1} 
}

\makeatletter
\newcommand*\ifthispageodd{%
  \checkoddpage
  \ifoddpage
    \expandafter\@firstoftwo
  \else
    \expandafter\@secondoftwo
  \fi
}
\makeatother

\usepackage[noindentafter]{titlesec} %Removes the indentation after new section
\counterwithout{section}{chapter} % Include if sections should be taken at the top level
\counterwithout{figure}{chapter} %Removes chapter number on figures
\counterwithout{table}{chapter} %Removes chapter number on tables
\numberwithin{equation}{section} % Sets numbering at the section level
\setsecnumdepth{subsubsection}

\DeclareMathAlphabet{\mathsfit}{OT1}{lmss}{m}{sl}
\DeclareMathAlphabet{\mathsfbf}{OT1}{lmss}{bx}{n}
\DeclareMathAlphabet{\mathsfbfit}{OT1}{lmss}{bx}{sl}
 
\DeclareMathVersion{chaptermath}
\SetSymbolFont{operators}{chaptermath}{OT1}{cmbr}{m}{n}
\SetSymbolFont{letters}{chaptermath}{OML}{cmbrm}{b}{it}
\SetSymbolFont{symbols}{chaptermath}{OMS}{cmbrs}{m}{n}
\DeclareMathVersion{subsectionmath}
\SetSymbolFont{operators}{subsectionmath}{OT1}{cmbr}{m}{n}
\SetSymbolFont{letters}{subsectionmath}{OML}{cmbrm}{m}{it}
\SetSymbolFont{symbols}{subsectionmath}{OMS}{cmbrs}{m}{n}

\titleformat{\section}{\centering \Large \bfseries \sffamily \mathversion{chaptermath} \color{blue!90!black} }{\thesection}{15pt}{}{}
\titlespacing{\section}{0pt}{15pt}{5pt}
\titleformat{\subsection}{\large \sffamily \mathversion{subsectionmath} \color{blue!90!black} }{\thesubsection}{10pt}{}{}
\titlespacing{\subsection}{0pt}{10pt}{5pt}
\titleformat{\subsubsection}{\normalsize \sffamily \itshape \mathversion{subsectionmath} \color{blue!80!black} }{\thesubsubsection}{10pt}{}{}
\titlespacing{\subsubsection}{0pt}{10pt}{0pt}

\newcommand{\sectionlike}[1]{\phantomsection \addcontentsline{toc}{section}{#1} \sectionmark{#1}
		\begin{center}
		\needspace{8\baselineskip}
		\Large \bfseries \sffamily \mathversion{chaptermath} \color{blue!90!black} #1  
		\end{center}
	\vspace{-5pt} 
}

\ifnum\fullheadfoot=1
	\makepagestyle{standardstyle}
	\makeoddhead{standardstyle}{\sffamily \mathversion{subsectionmath} \rightmark}{}{\sffamily \bfseries{\thepage}}
	\makeevenhead{standardstyle}{\sffamily \bfseries{\thepage}}{}{\sffamily \mathversion{subsectionmath} \leftmark}
	\makeheadrule{standardstyle}{\textwidth}{\normalrulethickness}
	\makepsmarks{standardstyle}{
		\nouppercaseheads
		\createmark{section}{both}{shownumber}{\ }{\hspace{1.2em}  }
		\createmark{subsection}{right}{shownumber}{}{\hspace{1.2em} }
	}
	
	\copypagestyle{appstyle}{standardstyle}
	\makeevenhead{appstyle}{\sffamily \bfseries{\thepage}}{}{ \sffamily \mathversion{subsectionmath} \leftmark}
	\makepsmarks{appstyle}{
		\nouppercaseheads
		\createmark{section}{both}{nonumber}{\ }{\ \ }
		\createmark{subsection}{right}{shownumber}{}{\ \ }
	}
\else %Plain style
	\makepagestyle{standardstyle}
	\makeoddfoot{standardstyle}{}{\sffamily -- {\bfseries\thepage} --}{} 
	\makeevenfoot{standardstyle}{}{\sffamily -- {\bfseries\thepage} --}{}
	\copypagestyle{appstyle}{standardstyle}
\fi

\createplainmark{toc}{both}{\contentsname}
\createplainmark{bib}{both}{\bibname}

\makeatletter
\let\MyIntOrig\int
\def\MyIntSpace{\hspace{-.35em}} %% Configure as needed.
\def\int{\MyInt}
\def\MyInt{\MyIntOrig\MyIntSkipMaybe}
\def\MyIntSkipMaybe{
	\@ifnextchar_{\MyIntSkipScript}{%
		\@ifnextchar^{\MyIntSkipScript}{%
			\@ifnextchar\limits{\MyIntSkipTok}{%
				\@ifnextchar\nolimits{\MyIntSkipTok}{%
					\MyIntSpace}}}}%
}
\def\MyIntSkipScript#1#2{#1{#2}\MyIntSkipMaybe}
\def\MyIntSkipTok#1{#1\MyIntSkipMaybe}

\newcommand{\pushright}[1]{\ifmeasuring@#1\else\omit\hfill$\displaystyle#1$\fi\ignorespaces}
\makeatother